\documentclass[trackchanges,twocolumn,twocolappendix]{aastex701}
\usepackage{amsmath}
\usepackage{CJKutf8}
\usepackage{xcolor}
\usepackage[T1]{fontenc}

\definecolor{mybrickred}{HTML}{B22222} 
\newcommand{\Will}{\color{black}} 

\begin{document}

\title{Chemical Abundances Shape History (CASH). I. A Link between Giant Planets Orbital Periods and Host Stellar C/O Ratios}

\author[orcid=0000-0001-5487-9450,gname=Ruisheng, sname='Zhang']{Ruisheng Zhang}
\affiliation{School of Astronomy and Space Science, Nanjing University, Nanjing 210093, China.}
\affiliation{Key Laboratory of Modern Astronomy and Astrophysics, Ministry of Education, Nanjing 210023, China.} 
\email{rszhang@smail.nju.edu.cn}

\author[orcid=0000-0002-6472-5348,gname=Ji-Wei, sname='Xie']{Ji-Wei Xie} 
\affiliation{School of Astronomy and Space Science, Nanjing University, Nanjing 210093, China.}
\affiliation{Key Laboratory of Modern Astronomy and Astrophysics, Ministry of Education, Nanjing 210023, China.} 
\email[show]{jwxie@nju.edu.cn}

\author[orcid=0000-0002-0162-163X,gname=Mengrui, sname='Pan']{Mengrui Pan} 
\affiliation{Institute for Astronomy, School of Physics, Zhejiang University, Hangzhou 310027, PRChina.}
\affiliation{Center for Cosmology and Computational Astrophysics, Institute for Advanced Study in Physics, Zhejiang University, Hangzhou 310027, PRChina.} 
\affiliation{Department of Earth, Environmental and Planetary Sciences, MS 126, Rice University, Houston, TX 77005, USA} 
\affiliation{Rice Space Institute, MS 126, Rice University, Houston, TX 77005, USA.} 
\email{mengruipan99@gmail.com}

\author[orcid=0000-0001-5830-3619,gname=Beibei, sname='Liu']{Beibei Liu} 
\affiliation{Institute for Astronomy, School of Physics, Zhejiang University, Hangzhou 310027, PRChina.}
\affiliation{Center for Cosmology and Computational Astrophysics, Institute for Advanced Study in Physics, Zhejiang University, Hangzhou 310027, PRChina.} 
\email{bbliu@zju.edu.cn}

\author[orcid=0000-0003-1680-2940,gname=Ji-Lin, sname='Zhou']{Ji-Lin Zhou} 
\affiliation{School of Astronomy and Space Science, Nanjing University, Nanjing 210093, China.}
\affiliation{Key Laboratory of Modern Astronomy and Astrophysics, Ministry of Education, Nanjing 210023, China.} 
\email{zhoujl@nju.edu.cn}

\author[orcid=0000-0002-4361-8885,gname=Ji, sname='Wang']{Ji Wang}
\affiliation{Department of Astronomy, The Ohio State University, 100 W. 18th Ave., Columbus, OH 43210, USA.}
\email{wang.12220@osu.edu}

\author[orcid=0000-0001-8618-3343,gname=Haiyang S., sname='Wang']{Haiyang S. Wang} 
\affiliation{Center for Star and Planet Formation, Globe Institute, the University of Copenhagen, Øster Voldgade 5–7, 1350 København K, Denmark.}
\email{haiyang.wang@sund.ku.dk}

\author[orcid=0000-0003-0097-4414,gname=Yapeng, sname='Zhang']{Yapeng Zhang} 
\affiliation{Department of Astronomy, California Institute of Technology, Pasadena, CA 91125, USA.}
\email{yapzhang@caltech.edu}

\begin{abstract}
The chemical abundance of host stars plays a pivotal role in shaping the formation history of planetary systems, yet the influence of elements beyond iron remains poorly understood. Here, we investigate the relationship between the carbon-to-oxygen (C/O) ratio of host stars and the orbital periods of giant planets. By analyzing high-resolution spectroscopic data from 598 planet-hosting stars (hosting 929 planets) across SDSS, Keck, and HARPS surveys, we identify a correlation: stars with higher C/O ratios are more likely to host longer-period giant planets. Theoretical models of pebble-driven planet formation and migration further support this observation, demonstrating that elevated C/O ratios enhance solid material availability at outer disk regions, promoting giant planet formation at larger distances and subsequent moderate inward migration. Our findings establish stellar C/O as a critical factor in shaping the orbital architecture of giant planets, bridging disk chemistry to planetary system evolution.
\end{abstract}

\keywords{\uat{Exoplanet formation}{492} --- \uat{Stellar abundances}{1577} --- \uat{Planet hosting stars}{1242} --- \uat{Planetary migration}{2206} --- \uat{Extrasolar gaseous giant planets}{509}}



\section{Introduction} 
{The intricate relationship between the properties of planets and the chemical element abundances of their host stars stands as a cornerstone in our quest to comprehend the history of planet formation and evolution} \citep{Teske2024ARA&A,2021ARA&A..59..291Z}. 
Prior investigations have predominantly centered on the iron abundance relative to hydrogen (Fe/H), which has illuminated critical aspects of planet formation 
\citep[e.g.][; for a review, see \citealt{Teske2024ARA&A}]{2001A&A...373.1019S,2005ApJ...622.1102F,2018PNAS..115..266D,2015AJ....149...14W,2019Geosc...9..105A}.

{\Will 
While Fe/H provides valuable insights, it alone is insufficient to fully elucidate the formation history of giant planets. 
Attention must also be directed towards other elements, particularly carbon (C), oxygen (O), and their ratio (C/O). 
These elements and their ratio are expected to play a pivotal role in shaping the chemical boundary conditions of planet formation. 
Carbon and oxygen are the two most abundant heavy elements in protoplanetary disks and dominate the major volatile carriers (e.g., $\mathrm{H_2O}$, CO, CO$_2$, and organics). 
Consequently, the stellar (and plausibly disk) C and O budgets---and especially their ratio C/O---help set the disk condensation sequence and regulate how volatiles partition between gas and solids across key snowlines, thereby modulating where solid mass is available for planetesimal growth and core assembly \citep{Turrini2021}. 
In turn, C/O is expected to imprint itself on (i) planetary bulk composition and internal structure, through the relative availability of oxygen- versus carbon-bearing condensates during assembly \citep{2010ApJ...715.1050B,2005astro.ph..4214K,2022MNRAS.513.5829W, Zaveri2026}, and (ii) atmospheric elemental and isotopic inventories, because forming and migrating planets can accrete gas and solids from chemically distinct reservoirs \citep{Oberg2011,2023AJ....165....4W,2024AJ....168..246Z, Franco2026}. 
Although disk chemistry and transport are complex and can cause local deviations from stellar values \citep{Oberg2023ARA&A}, stellar photospheric abundances nonetheless provide an empirically accessible proxy for the natal composition of planet-forming material \citep{2025TrGeo...701194H, Teske2024ARA&A}. This motivates the expectation that stellar C/O may correlate with planet properties in a statistical sense even if not deterministically for individual systems.
However, the nature of the relationship between stellar C/O ratios and planetary systems remains poorly constrained \citep{Teske2024ARA&A}, with a particular lack of observational evidence and coherent understanding regarding whether and how the stellar carbon, oxygen, and the C/O ratio influence planetary system architectures \citep{2011ApJ...735...41P, 2014A&A...568A..25N, 2018A&A...614A..84S, 2021MNRAS.504.4252M, 2022ApJS..259...45T,Baburaj2026}.}

{ \Will
In this work, we investigate the role of the host star's C/O ratio from a novel perspective—the final orbital period of the giant planet. 
Specifically, we test the hypothesis that the host star's initial C/O ratio systematically influences the migration pathways and final orbital architectures of giant planets by dictating the solid surface density profile around major ice lines. 
The rationale behind this hypothesis is that varying carbon and oxygen abundances alter the availability of water ice and carbonaceous grains. This, in turn, affects the local solid mass available for rapid core growth, shifts the optimal locations for the onset of runaway gas accretion, and ultimately modifies the extent of disk-driven planetary migration \citep{2017A&A...602A..21S,2021PhR...893....1O,2025A&A...701A..47X,2021A&A...647A..96B}. 
By observationally linking stellar C/O ratios to giant planet orbital periods, we aim to uncover the imprint of these formation and migration mechanisms, establishing an intrinsic connection between primordial stellar chemistry and planetary system properties.}

\section{Sample Selection}

We constructed a {\Will unified catalog} of planet hosts by combining high-resolution spectroscopic data from three major surveys: SDSS-IV/APOGEE DR17 \citep{2017AJ....154...28B, 2022ApJS..259...35A}, Keck (CPS and CKS; \citealt{2016ApJS..225...32B, 2018ApJS..237...38B}), and HARPS \citep{2021A&A...655A..99D}. 
To ensure reliable stellar parameters and minimize evolutionary effects on planetary architectures, we restricted our sample to main-sequence stars with $5000 \le T_{\mathrm{eff}} \le 6500$\,K and $\log g > 4$.

For the APOGEE DR17 sample ($\sim$72,000 stars), we utilized parameters from the ASPCAP pipeline and applied standard quality cuts, requiring a signal-to-noise ratio (SNR) $> 80$ and excluding stars with problematic flags (e.g., \texttt{STAR\_BAD}, \texttt{METALS\_BAD}). To mitigate large uncertainties in volatile elements, we further required abundance errors in [C/H] and [O/H] to be $< 0.1$\,dex.
For the Keck sample, we combined data from the California Planet Search (CPS) and California-Kepler Survey (CKS). To maintain abundance precision consistent with APOGEE ($\sigma_{\rm [X/H]} \sim 0.03$--$0.04$\,dex), we adopted a stricter threshold of SNR $> 120$.
Similarly, for the HARPS sample, we selected stars from the GTO program with SNR $> 300$, yielding comparable measurement uncertainties ($\sim 0.02$\,dex for C and $\sim 0.05$\,dex for O).

We cross-matched these stellar samples with confirmed planet hosts from the NASA Exoplanet Archive (accessed 2025 May 29, \citealt{2025arXiv250603299C}). To eliminate duplicates among the surveys, we employed a hierarchical selection scheme based on sample size and homogeneity: SDSS-IV measurements were prioritized, followed by Keck, and finally HARPS. 
We verified that the main C/O trends persist within each individual survey subsample, confirming that our results are not artifacts of this specific prioritization scheme (see Section \ref{subsec:survey_heterogeneity}).
This procedure yielded a final {\Will unified catalog} of 598 unique host stars harboring 929 confirmed planets.
While most planets in our sample have direct radius measurements (e.g., from Kepler), 117 planets discovered by other methods possess only mass constraints. {\Will For these targets, we adopt radii estimated via the probabilistic mass-radius relation of \citet{2017ApJ...834...17C}. We have verified that our results are robust to this choice by cross-checking with the more recent relation of \citet{Otegi2020}, finding no significant change in our conclusions.}

\section{Results} 
\subsection{Period - C/O correlation from Observation}
\subsubsection{C/O distribution: Hot vs. Cold}

Leveraging this unified catalog, we investigate the relationship between host star C/O ratios and planetary orbital periods.

\begin{figure*}[htbp]
\centering
\includegraphics[scale = 0.43]{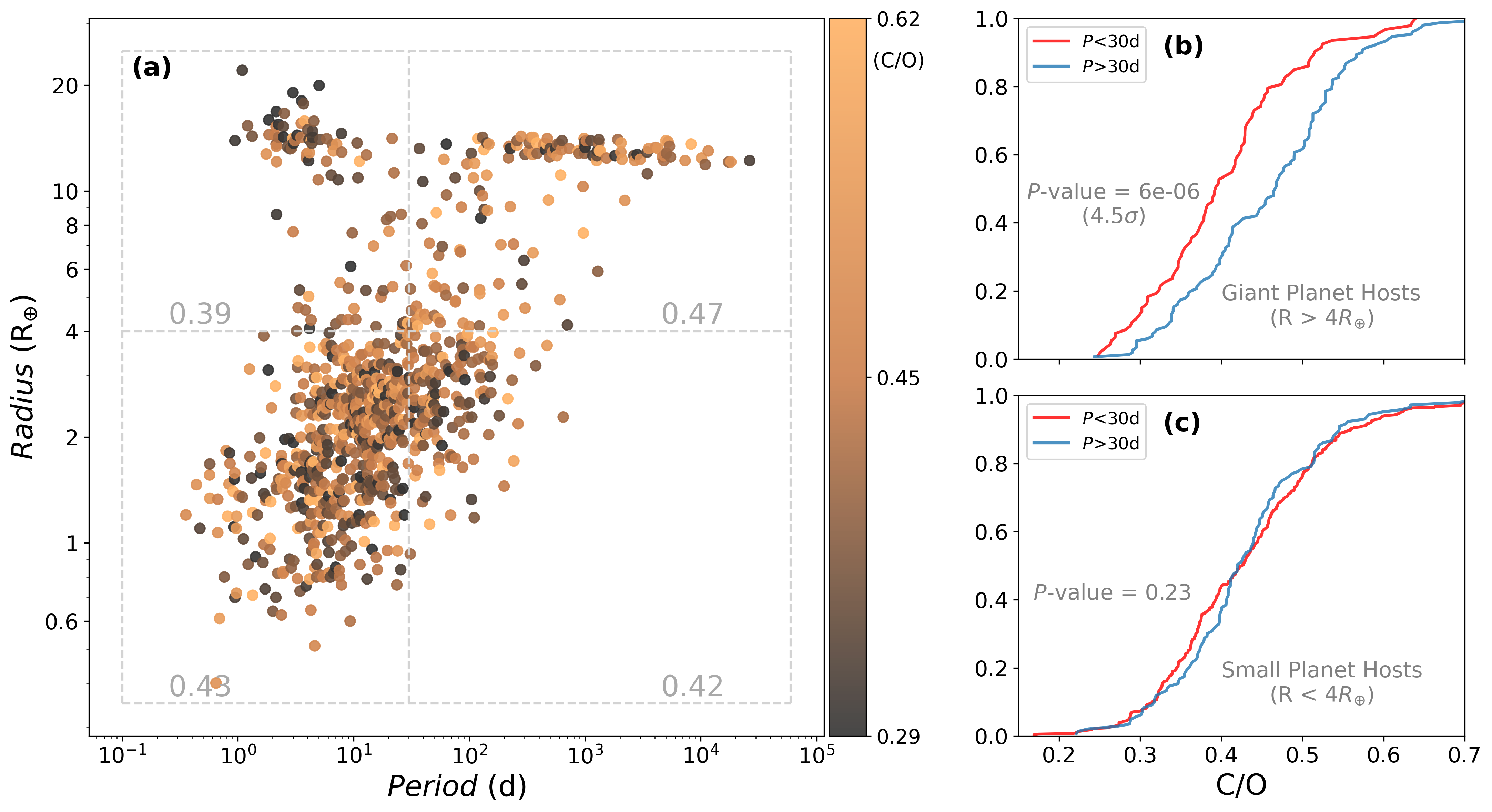}

\caption{\textbf{Period-Radius and C/O Ratio Distribution in Hosts with Different Types of Planets.}
    \textbf{Panel (a)} is the period-radius diagram of planets, with the C/O ratio of their host stars color-coded. The dashed lines ($P = 30 $d, $Radius=4R_{\oplus}$) divide the figure into four parts: the top two parts are hot/cold giant planets, while the bottom panels are hot/cold small planets. The gray numbers in each part indicate the median C/O ratio of the respective regions. 
    \textbf{Panel (b)} compares the cumulative distributions of the C/O ratios for stars hosting giant planets with short ($<30$d) and long ($>30$d) orbital periods and radii $R_{\rm{p}}$ $>$ 4$R_{\oplus}$.
    \textbf{Panel (c)} is analogous to panel (b), but it focuses on small planets with $R_{\rm{p}}$ $<$ 4$R_{\oplus}$.} 

\label{fig:Figure1}
\end{figure*}

Figure \ref{fig:Figure1}(a) displays the stellar C/O distribution in the planetary period-radius diagram. As observed, cold giant planets ($R_{\rm p} > 4 R_{\oplus}$) with long orbital periods ($P > 30$ days) tend to have higher host star C/O ratios (median C/O = 0.47) compared to hot giant planets with short periods (median C/O = 0.39). In contrast, small planets ($R_{\rm p} < 4 R_\oplus$) generally exhibit similar host-star C/O ratios irrespective of their orbital periods, with median values of 0.43 for short-period systems and 0.42 for long-period systems. The results presented in Figure \ref{fig:Figure1}(a) are further supported by Figures \ref{fig:Figure1}(b) and \ref{fig:Figure1}(c), which compare the cumulative distributions of host star C/O between long-period and short-period planets for both giant planets and small planets, respectively. Specifically, for giant planets, the C/O distribution of hosts with long-period planets differs significantly from that of short-period planet hosts, with a Kolmogorov-Smirnov (K-S) test $P$-value of $6\times 10^{-6}$. Conversely, the two C/O distributions are comparable for small planets, yielding a K-S test $P$-value of 0.23. {\Will We note that when the analysis is repeated on individual survey subsamples, the significance is reduced (to 3.4$\sigma$ for SDSS-IV, 2.3$\sigma$ for HARPS, ~1.4$\sigma$ for Keck; see Section \ref{subsec:survey_heterogeneity}), as expected given the smaller sample sizes. Similarly, inter-survey abundance calibration can modestly affect the inferred significance (Section \ref{subsec:abundance_calibration}). }

\begin{figure*}[t!]
\centering
\includegraphics[scale = 0.4]{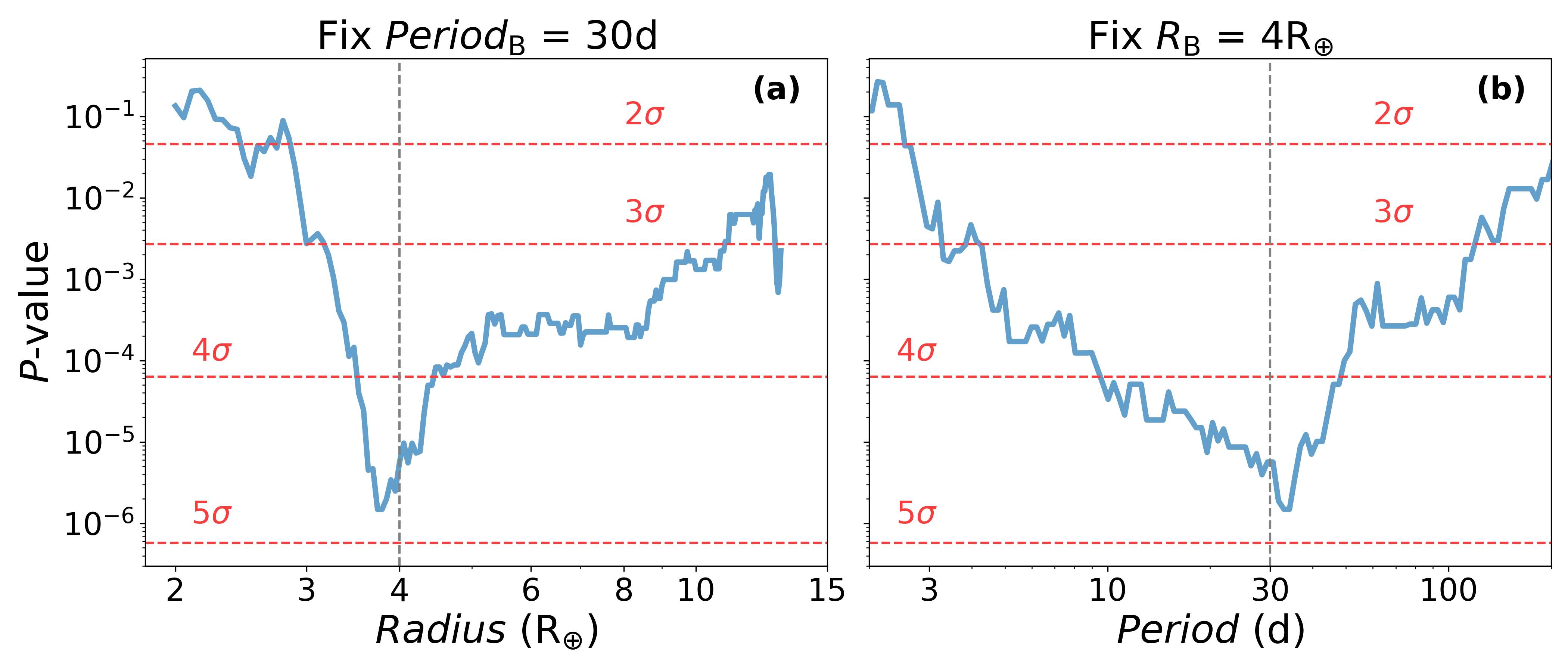}
\caption{\textbf{\textit{P}-values from K-S tests Comparing C/O Distributions of Exoplanet Hosts across Radius (a) and Period (b) Selection.}
    This figure shows the \textit{P}-values from Kolmogorov–Smirnov (K-S) tests, which assess whether the C/O distributions of short-period and long-period exoplanets are drawn from the same parent population. 
    \textbf{Panel (a)} shows the \textit{P}-value as a function of the chosen radius boundary for selecting giant planets, with the orbital period boundary fixed at 30 days. The vertical dashed line marks the fiducial radius boundary at $R_{\rm{B}} = 4R_\oplus$. 
    \textbf{Panel (b)} shows the \textit{P}-value as a function of the chosen period boundary for separating short/long-period giant planets, while fixing the radius boundary at $R_{\rm{B}} = 4R_\oplus$. The vertical dashed line marks the fiducial period boundary at 30 days. 
    }
\label{fig:Figure_2_pvalmap}
\end{figure*}

To investigate how the above results depend on the radius and period boundaries ($R_{\rm p} = 4 R_{\oplus}$ and $P = 30$ days, as adopted in Figure \ref{fig:Figure1}), we repeated the K-S test by varying these boundaries. 
The corresponding K–S test results are presented in Figure~\ref{fig:Figure_2_pvalmap}(a) and \ref{fig:Figure_2_pvalmap}(b), which display the variation of the K–S test $P$-value with the adopted radius and period boundaries, respectively.
As can be seen, our results are not highly sensitive to specific boundary values; the $P$-value remains significant ($<10^{-3}$, corresponding to a confidence level of $>3\sigma$) across a wide range of radius and period boundaries. 
However, the period boundary around $P = 30$\,d and the radius boundary around $R_{\rm p} = 4 R_{\oplus}$ yield the most significant results with the smallest K-S test $P$-values.
Interestingly, $R_{\rm p} = 4 R_{\oplus}$ generally coincides with the radius “cliff” observed in the occurrence rate distribution of exoplanets \citep{2023AJ....166..122D}, and this value is a commonly used threshold to distinguish gas giant planets from smaller planets \citep{2021ARA&A..59..291Z}. 
{Similarly, a period of 30 days ($P = 30\,\mathrm{d}$) lies at the heart of the transitional region that characterizes warm Jupiters—a class of giant planets that bridge the gap between hot and cold Jupiters \citep{2018ARA&A..56..175D}.}

Panel (a) illustrates the sensitivity of the $P$-value to the radius threshold, with the period boundary fixed at $P_\mathrm{B}=30$\,days. {\Will The statistical significance is maximized (i.e., minimum $P$-value) near $R_\mathrm{B} \approx 3.8\,R_\oplus$, close to the commonly adopted threshold of $4\,R_\oplus$, confirming that this boundary provides the optimal separation in host-star C/O distributions between small and giant planets. }

Complementarily, Panel (b) displays the results when fixing the radius boundary at $R_\mathrm{B}=4\,R_\oplus$ and varying the period cut. {The strongest contrast between short- and long-period systems emerges near $P_\mathrm{B} \approx 35$\,days, close to our fiducial boundary of 30\,days and consistent with the transition region between hot and warm Jupiters.}

These parameter sweeps demonstrate that the C/O–period relation is robust against moderate variations in boundary definitions. The persistence of the trend over a broad parameter space indicates that the observed C/O dichotomy reflects a genuine astrophysical distinction rather than an artifact of fine-tuned selection thresholds.

\begin{figure}
\centering
\includegraphics[width=\columnwidth]{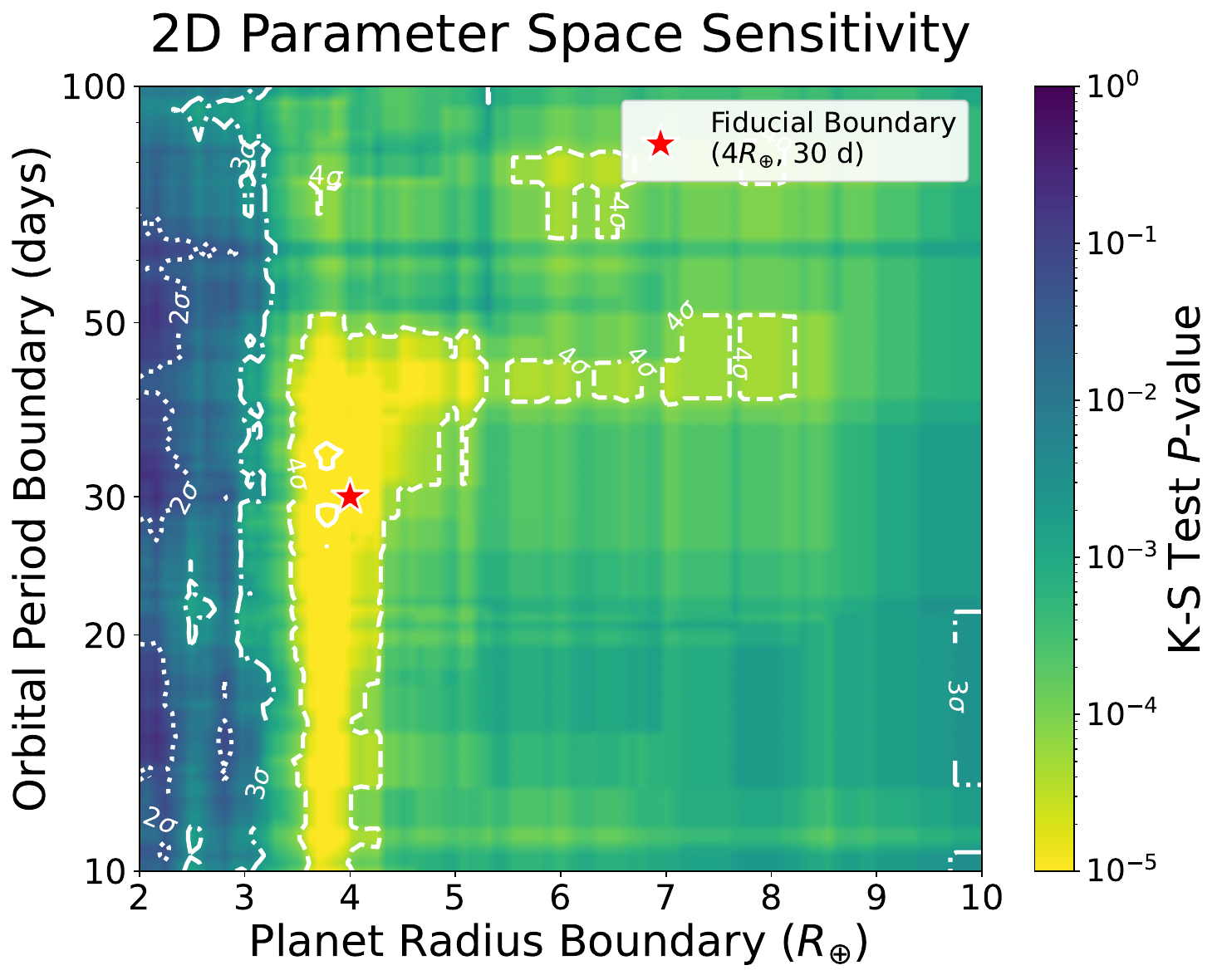}
\caption{Two-dimensional parameter space sensitivity of the K-S test.  
The color scale shows the K-S test $P$-value comparing the C/O distributions 
of short- and long-period giant planet hosts as both the planet radius boundary 
($x$-axis) and the orbital period boundary ($y$-axis) are varied simultaneously.  
White contours mark the 2$\sigma$, 3$\sigma$, and 4$\sigma$ significance levels.  
The red star indicates the fiducial boundary ($4\,R_\oplus$, 30\,d).  
The region of high statistical significance is broad and contiguous, spanning 
$R_\mathrm{B} \approx 3$--$5\,R_\oplus$ and $P_\mathrm{B} \approx 10$--$45$\,d, 
with a sharper peak along the radius axis than the period axis.}
\label{fig:2d_sensitivity}
\end{figure}

{\Will
To visualize this more clearly, Figure~\ref{fig:2d_sensitivity} presents a two-dimensional sweep over the radius and period boundaries simultaneously. The region of high statistical significance is broad and contiguous, spanning approximately $R_\mathrm{B} \approx 3$–$5 R_\oplus$ in radius and $P_\mathrm{B} \approx 10$–$45$ days in period. Importantly, the sensitivity is more sharply peaked along the radius axis, with a clear optimum near $R_\mathrm{B} \approx 3.8 R_\oplus$, whereas along the period axis the $P$-value remains low across a substantially wider interval. This pattern indicates that the radius boundary is more tightly constrained than the period boundary, and supports the interpretation that the C/O signal is primarily associated with a continuous dependence on orbital period rather than a transition at a uniquely special period. In this context, our fiducial choice of $P = 30$ days should be viewed not as a singularly preferred boundary, but as a representative cut within the broader significant range. It remains a practical and physically motivated choice because it lies near the conventional hot/warm Jupiter divide \citep{2018ARA&A..56..175D}, while the main conclusions do not depend sensitively on the exact period threshold adopted.}

We note that all planetary parameters are drawn from the NASA Exoplanet Archive's Planetary Systems Composite Table \citep{2025arXiv250603299C}, which provides a unified default parameter set for each planet by prioritizing the most precise peer-reviewed measurements. This avoids subjective selection 
among discrepant literature values. Moreover, typical uncertainties in period and radius are too small to move most planets across the adopted classification boundaries, and the broad stability of the 2D parameter sweep confirms that even if a few planets were to shift between groups, our conclusions would remain unchanged. 

\subsubsection{Period-C/O Correlation Analyses}

\begin{figure*}[htbp]
\centering
\includegraphics[width=0.95\textwidth]{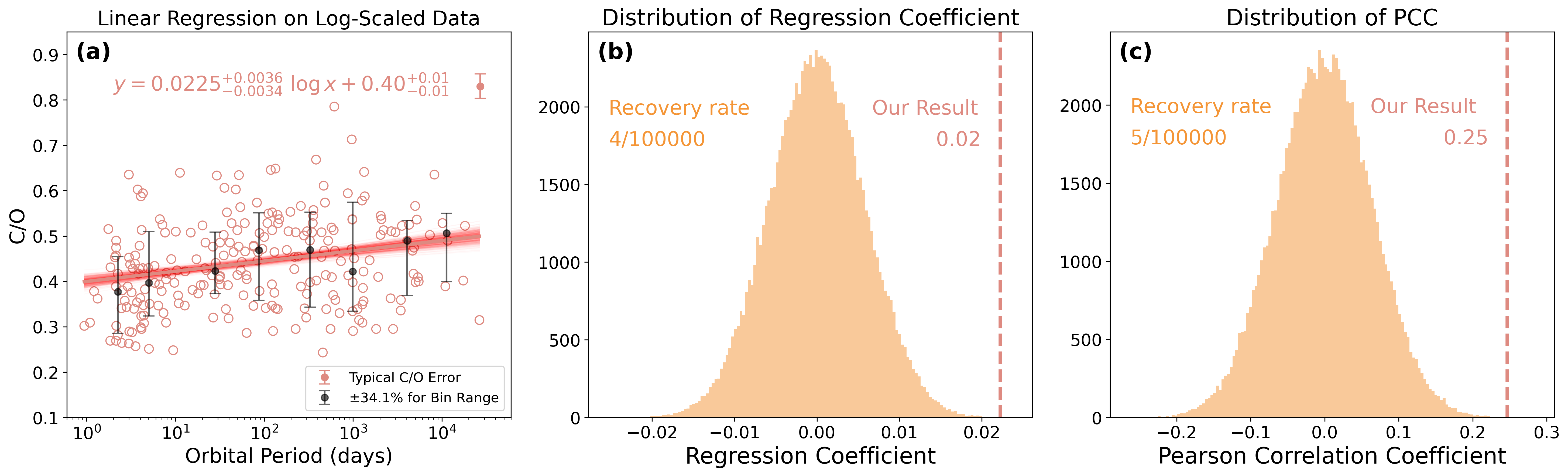}
\caption{\textbf{The Positive Correlation between Giant Planet Orbital Periods and Stellar C/O Ratios.}
    \textbf{Panel (a)} shows the result of a linear regression on log-scaled orbital period data, revealing a positive slope in C/O with increasing period. \textbf{Panels (b) and (c)} display the distributions of regression coefficients and Pearson correlation coefficients (PCC), respectively, from 100,000 random permutations of the data. The vertical dashed lines indicate the observed values.
}
\label{fig:Figure3}
\end{figure*}

We further investigate the relation between the C/O of host stars and the periods of {243 giant planets in Figure \ref{fig:Figure3}}. 
Specifically, panel (a) of Figure \ref{fig:Figure3} displays C/O as a function of period for giant planets in our sample. 
{Each red circle represents an individual data point. The red vertical error bar on the top-right indicates the typical uncertainty in the C/O measurement.
The black error bars reflect the 50$\pm$34.1 percentiles of C/O in 8 period bins. Although there is considerable scatter, a significant correlation is still observed with a Pearson correlation coefficient of 0.27 and a corresponding nominal $P$-value of 0.0001.
We employ a simple linear model to fit the correlation in the C/O-$\log(\text{period})$ plane, obtaining: 
\begin{equation}
\mathrm{C/O}= \left(0.0225^{+0.0036}_{-0.0034}\right) \cdot \log(\text{period}) + \left(0.40^{+0.01}_{-0.01}\right)
\label{eq:CO_period}
\end{equation}

{To account for measurement uncertainties in the C/O ratio, we estimated the uncertainties of the fitting parameters through a resampling–refitting procedure. Specifically, for each data point, the observed [C/H] and [O/H] values were randomly resampled according to their reported uncertainties, assuming normal error distributions. The resampled [C/H] and [O/H] values were then used to recompute the corresponding C/O ratios, which were subsequently refitted to derive a new best-fit relation. This process was repeated 100,000 times, and the resulting ensemble of best-fit lines forms the shaded red region, representing the propagated uncertainty in the fit due to measurement errors.}

To further quantify the significance of this correlation, we also conducted a bootstrap test as follows. First, we constructed a bootstrap trial sample by shuffling the C/O and period values in the observed sample to create random pairings. {The sample size is 243, which is the same sample used in the Pearson test.} We then calculated the Pearson correlation coefficient and fit the same linear model to the bootstrap trial sample. In total, 100,000 bootstrap trial samples were generated and analyzed, and the distributions of their Pearson correlation coefficients and the linear fitting slopes are plotted in the right and middle panels of Figure \ref{fig:Figure3}, respectively. As can be seen, only 4(5) out of 100,000 trials resulted in a Pearson correlation coefficient (linear fitting slope) larger than the observed values, translating to a significance $P$-value of $4(5)\times 10^{-5}$ and corresponding to a confidence level of greater than $4\sigma$.

To further assess the statistical connection between stellar C/O ratios and giant planet orbital periods, we compared two models: a constant model ($M_0$) representing no trend, and a log-linear model ($M_1$) that allows for a positive trend of C/O with the logarithm of orbital period.

\begin{figure*}[htbp]
    \centering
    \includegraphics[width=0.95\textwidth]{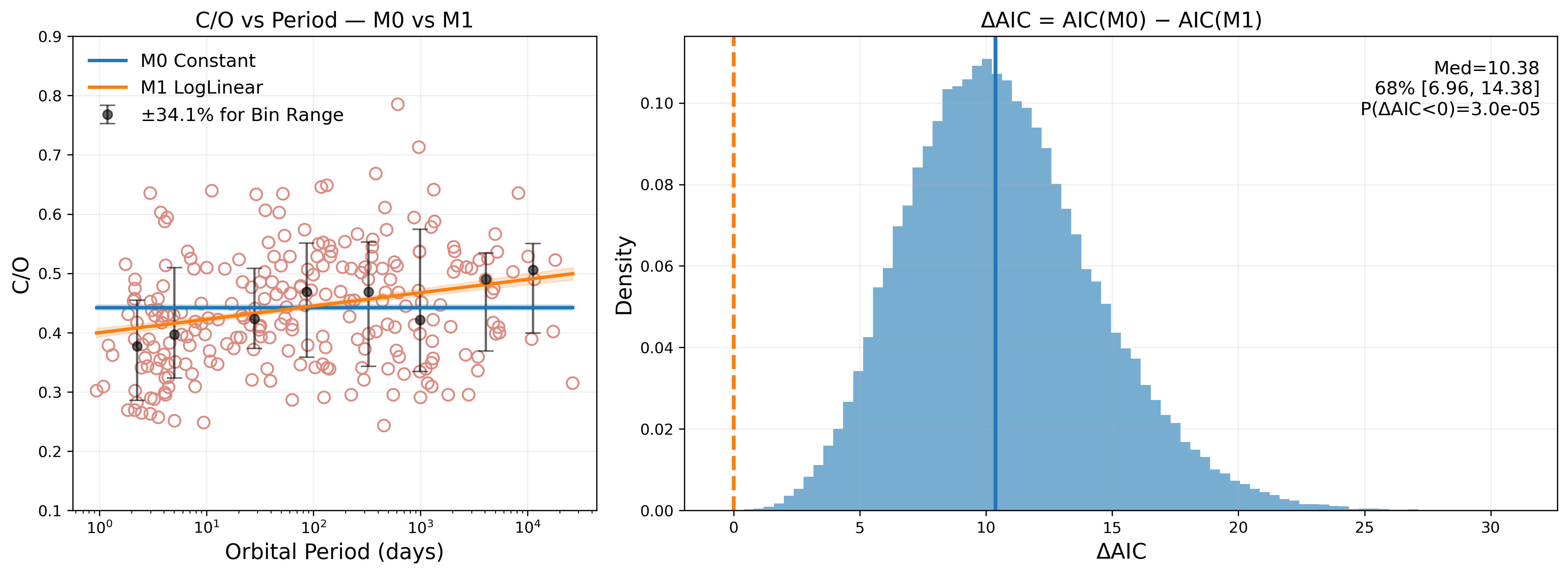}
    \caption{
        \textbf{C/O ratio versus orbital period and model comparison.}
        \textbf{Left:} Monte Carlo–averaged fits comparing two models for the planetary C/O ratio as a function of orbital period:
        a constant model ($M_0$; blue) and a log–linear model ($M_1$; orange).
        Each curve shows the median fit with shaded $\pm1\sigma$ bands across $10^5$ trials, 
        and the black points mark the median and $\pm34.1\%$ range within logarithmic period bins.
        \textbf{Right:} Distribution of $\Delta\mathrm{AIC} = \mathrm{AIC}(M_0) - \mathrm{AIC}(M_1)$ 
        from $10^5$ Monte Carlo realizations, with the median (blue) and the null reference line at $\Delta\mathrm{AIC}=0$ (orange dashed).
        Positive $\Delta\mathrm{AIC}$ values indicate that the log–linear model provides a statistically better description of the data than the constant model.
        }
\label{fig:CO_models}
\end{figure*}

To propagate measurement uncertainties into both the fitted C/O--period relation and the model-selection statistics, we adopted a unified Monte Carlo resample--refit procedure with $10^5$ realizations. In each realization, we drew perturbed abundances $C_i^{(s)} \sim \mathcal{N}(C_i, \sigma_{C,i})$ and $O_i^{(s)} \sim \mathcal{N}(O_i, \sigma_{O,i})$ for every system (where $C_i, O_i$ are the measured values and $\sigma_{C,i}, \sigma_{O,i}$ their reported uncertainties), recomputed
\[
    \mathrm{C/O}_i^{(s)} = 10^{\,C_i^{(s)} - O_i^{(s)}}\,(\mathrm{C/O})_\odot,
\]
and independently refit a constant model ($M_0$; one free parameter) and a log--linear model ($M_1$; two free parameters). For each trial we recorded the resulting model curves and computed $\Delta\mathrm{AIC}^{(s)} = \mathrm{AIC}(M_0) - \mathrm{AIC}(M_1)$. This framework ensures that observational errors are consistently propagated into both the confidence envelopes of the fitted trends (Figure~\ref{fig:CO_models}, left) and the $\Delta\mathrm{AIC}$ distribution used for model selection (Figure~\ref{fig:CO_models}, right).

As shown in Figure~\ref{fig:CO_models}, left panel, the log-linear model (orange line) introduces a mild positive slope, suggesting that C/O tends to increase with orbital period. The $\pm 1\sigma$ confidence band derived from $10^5$ Monte Carlo realizations which considered observational errors indicates that this positive trend is statistically consistent across resamplings, although the scatter among individual systems remains significant. The constant model (blue line), in contrast, assumes a uniform mean C/O value and fails to capture the gradual rise observed toward longer-period planets.

The statistical comparison based on the Akaike Information Criterion (AIC) further supports this result. The right panel of Figure~\ref{fig:CO_models} shows the distribution of $\Delta\mathrm{AIC} = \mathrm{AIC(M_0)} - \mathrm{AIC(M_1)}$ obtained from $10^5$ Monte Carlo trials. The distribution peaks at $\Delta\mathrm{AIC} \approx 10$, with a median value of $10.38^{+3.99}_{-3.42}$ (68\% interval $[6.96, 14.38]$), only $P(\Delta\mathrm{AIC}<0)=3\times10^{-5}$ of the trials prefer the constant model. This indicates that the log-linear model provides a consistently better fit than the constant one, with very strong statistical support according to standard AIC significance thresholds ($\Delta\mathrm{AIC} > 10$).

In summary, both visual and information-theoretic evidence suggest that a statistically significant positive correlation exists between host-star C/O and orbital period among giant planets. The improvement of the log-linear model over the constant baseline implies that this correlation is unlikely to arise from random scatter alone, reinforcing the hypothesis that host-star C/O ratios are correlated with the orbital periods of their giant planets.

\subsection{Period - C/O Correlation from Simulation}

Since carbon and oxygen are the most abundant elements and key constituents of major molecular species in protoplanetary disks, their distribution and partitioning play a crucial role in determining the disk's chemical composition and planet formation processes. 
Carbon exists in various forms, such as refractory carbon grains, organic hydrocarbons ($\mathrm{C}_{x}\mathrm{H}_{y}$), carbon dioxide ($\mathrm{CO}_2$), and carbon monoxide (CO). 
Oxygen-bearing species include silicates and water ($\mathrm{H}_2\mathrm{O}$), as well as the aforementioned carbon-containing molecules $\mathrm{CO}_2$ and CO. 

\begin{figure}[htbp]
\centering
\includegraphics[scale = 0.15]{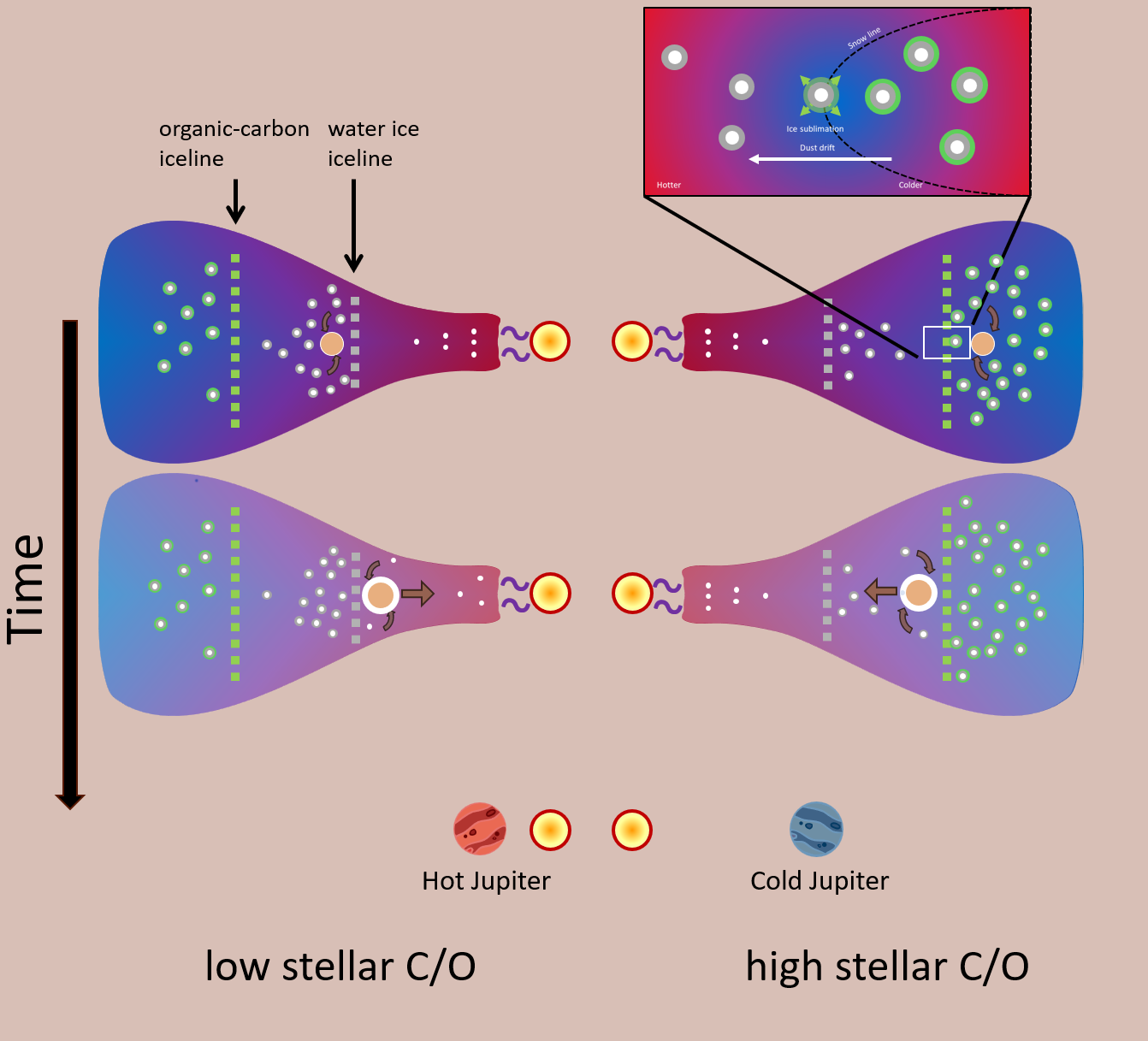}
    \caption{\textbf{Schematic illustration of how the C/O ratio of host stars regulate the formation, migration and final orbital periods of giant planets.}
        In the left panel (low stellar C/O case), the solid material is more concentrated in the inner oxygen-rich disk region, and the protoplanet is more likely to form near the water (H$_2$O) ice line via streaming instability. It subsequently undergoes inward migration by accreting solids and gas, ultimately forming a hot giant planet with a short orbital period. In the right panel (high stellar C/O case),  the protoplanet forms at the farther out disk region near the organic carbon and carbon dioxide ice line. After moderate inward disk migration, the planet ends up into a giant planet with a relatively long orbital period.
}
\label{fig:Figure_5}
\end{figure}

As depicted in Figure \ref{fig:Figure_5}, we elucidate how the stellar C/O ratio influences the formation and evolution of giant planet systems.
{Since stars and their protoplanetary disks originate from the same molecular cloud, the metallicity of the host star is conventionally considered a reasonable proxy for that of the natal protoplanetary disk.}
In oxygen-rich disks characterized by a lower C/O ratio, a large proportion of volatile oxygen can be sequestered in $\mathrm{H}_2\mathrm{O}$. 
In contrast, in carbon-rich disks with a higher C/O ratio, organic hydrocarbons and $\mathrm{CO}_2$ become the dominant species. The ice lines of these two carbon-rich species are situated farther out in the disk compared to the water ice line. As a result, solid material is more concentrated at the outer regions of disks with a higher C/O ratio. 

{\Will We emphasize that this picture is a theoretical prediction grounded in condensation chemistry and elemental mass balance 
\citep{2010ApJ...715.1050B, Oberg2011, Turrini2021}, rather than a directly 
observed phenomenon. The redistribution of solid mass toward outer ice 
lines at higher C/O follows as a straightforward consequence of the 
reduced oxygen budget available for H$_2$O formation. Recent ALMA 
observations lend indirect support: the MAPS Large Program 
\citep{Oberg2021, Law2021} has revealed radially varying molecular 
abundances consistent with snowline chemistry. Direct spatially resolved measurements of the solid mass 
distribution in disks with known stellar C/O ratios remain an important 
goal for future observations; our statistical link between stellar C/O 
and giant planet orbital periods provides a complementary, indirect test 
of this theoretical prediction.}

Ice lines are considered as the birthplaces of protoplanets. When the solid material in the form of mm-cm sized pebbles drifts across the ice line, their corresponding chemical compositions sublimate, enriching the surrounding vapor abundance. The vapor within the ice line can diffuse back and re-condense onto exterior icy pebbles. This process substantially increases the local solid-to-gas density ratio and initiates the formation of planetesimals and protoplanets through streaming instability \citep{2017A&A...602A..21S, 2017A&A...608A..92D}.

In disks with a higher C/O ratio, protoplanets predominantly form at the ice lines of carbon-rich species in the outer disk regions.
{As previously mentioned, the solid material is less concentrated in the inner regions of high-C/O disks, which further limits planetary growth and the range of inward migration.} Taken together, giant planets forming in disks with a higher C/O ratio tend to have longer orbital periods, while those forming in disks with a lower C/O ratio end up as hot/warm Jupiters.

\begin{figure}[htbp]
\centering
\includegraphics[scale = 0.4]{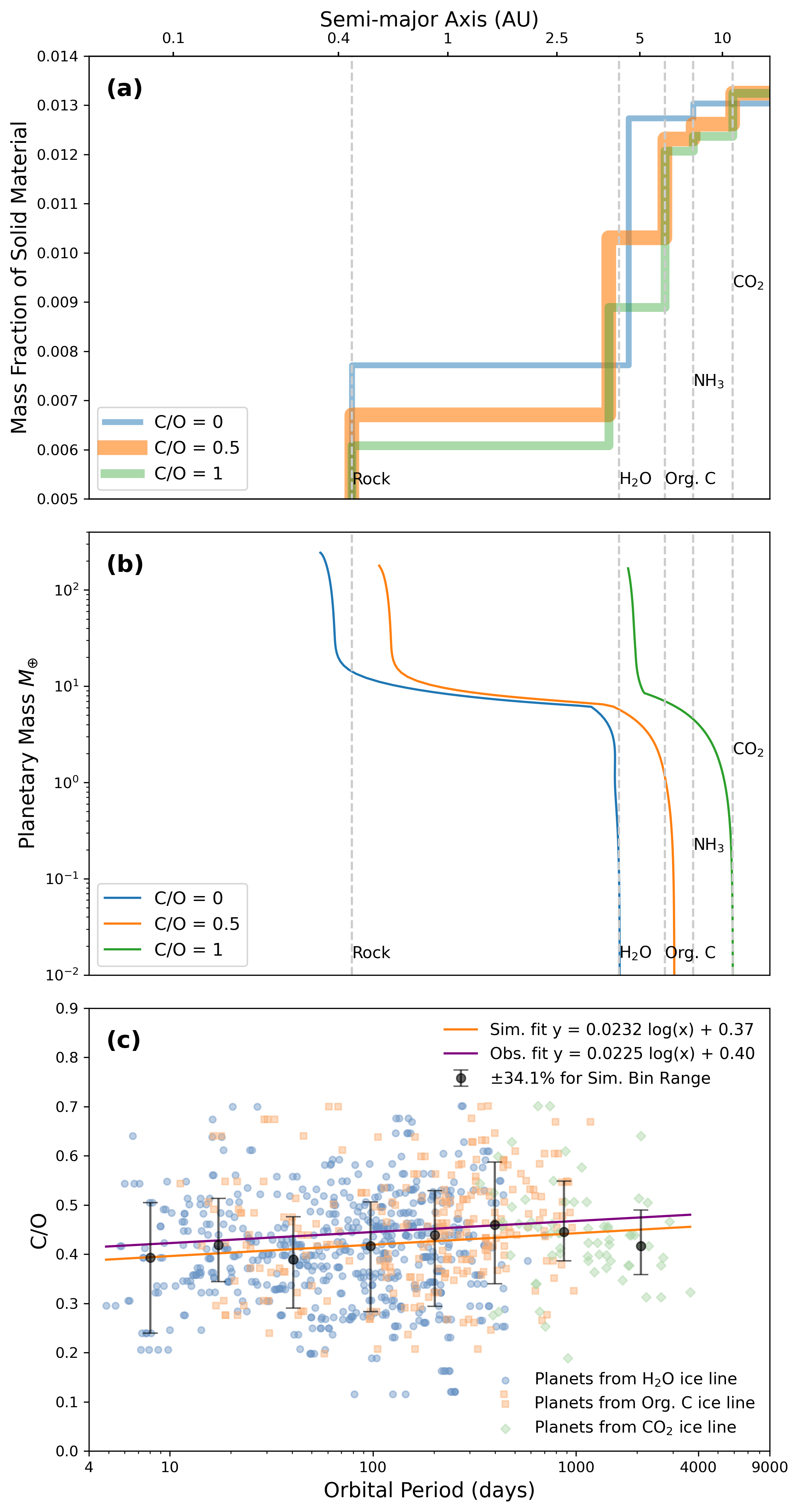}
    \caption{\textbf{Impact of stellar C/O ratio on solid material distribution, formation and migration of giant planet, and C/O–orbital period trend.}
            \textbf{Panel (a):} mass fraction of solid material as a function of radial distance from the central star with three characteristic stellar C/O of $0$, $0.5$, and $1$, respectively. The vertical dashed lines refer to the major ice lines of silicate, water, organic carbon, and ammonia and carbon dioxide from left to right.
            \textbf{Panel (b):} The growth and migration of giant planets where the birth protoplanets of $10^{-2} \ M_{\oplus}$ are initiated at different species of ice lines according to the stellar C/O ratios. 
            \textbf{Panel (c):} Stellar C/O versus planetary orbital period from population-synthesis models. Blue, orange, and green points represent planets formed near the H$_2$O, organic carbon, and CO$_2$ ice lines, respectively. The orange and purple lines denote the best-fit relations from simulations and observations.}
            
\label{fig:IR}
\end{figure}

{Figures \ref{fig:IR}(a) and \ref{fig:IR}(b) present the radial distribution of disk solids and the growth and migration of protoplanets with three characteristic stellar C/O ratios of $0$, $0.5$, and $1$, respectively. 
As can be seen, when the host star has a C/O ratio of $1$, the protoplanet forms in the outer disk region close to the ice line of carbon dioxide.
After formation, it undergoes disk migration towards the inner disk region characterized by lower solid densities (green line in Figure \ref{fig:IR}(a)), impeding the rapid growth rate and migration. Given the initial distant formation location and subsequent slow accretion, the protoplanet eventually grows into a gas giant planet with a relatively long orbital period of 1805 days. 
For comparison, when the star has a C/O ratio of $0$, solid material is more abundant in the inner disk region (blue line in Figure \ref{fig:IR}(a)). 
After formation around the $\mathrm{H_2O}$ ice line, the protoplanet accretes more solids and experiences shorter distance migration during the gas-rich disk phase, eventually ending up as a hot/warm Jupiter at a location close to its stellar host.}

To validate such a correlation quantitatively, we conduct a planet formation population synthesis model by Monte Carlo sampling the initial conditions (see Appendix \ref{sec:pop_synthesis} for details).
The giant planets (more massive than $30~M_{\oplus}$) are depicted in Figure \ref{fig:IR}(c).
The best linear fit for the planetary orbital period in relation to the stellar C/O ratio is $\mathrm{C/O}= 0.0232 \cdot \log(\text{period}) + 0.37$ (orange line), closely resembling the observed correlation (purple line).

{\Will 
We note that the three individual migration tracks shown in Figure~\ref{fig:IR}(b) represent only characteristic examples for each C/O value and are not intended to capture the full diversity of outcomes. Even for a fixed stellar C/O ratio, the final orbital period of a giant planet is subject to considerable stochastic variation due to differences in initial disk conditions, including the accretion rate, turbulent strength, and the precise birth location of the protoplanet. This intrinsic dispersion is clearly reflected in the full population synthesis results of Figure~\ref{fig:IR}(c), where the scatter among simulated planets is substantial even within a single formation channel. Notably, a significant number of simulated planets---particularly those formed near the H$_2$O ice line (blue points)---do migrate inward to orbital periods well inside 10~days, demonstrating that the pebble-driven framework is capable of producing short-period giants under favorable disk conditions.

Figure~\ref{fig:IR}(c) also reveals a systematic gradient across the three formation channels: planets originating at the CO$_2$ ice line (green points) and the organic carbon ice line (orange points), which are statistically associated with higher stellar C/O ratios, preferentially reside at longer final orbital periods compared to those formed near the H$_2$O ice line (blue points). Although the three populations partially overlap---some outer-forming planets do migrate to shorter periods---this overlap is asymmetric in probability. Planets born at the outer carbon-rich ice lines must traverse a substantially larger radial distance to reach ultra-short-period orbits before disk dissipation, making such extreme inward migration statistically less likely than for their counterparts starting at the H$_2$O ice line. This probabilistic asymmetry is why the initial chemical spatial gradient imposed by the stellar C/O ratio is statistically preserved in the final orbital architecture, despite population mixing.
}

The population synthesis model presented here serves primarily as an initial demonstration, confirming that the observed C/O–period correlation is a natural and logical outcome consistent with theoretical expectations. 
However, more detailed and comprehensive theoretical and numerical simulations in future work remain essential to provide crucial quantitative constraints on fundamental theoretical parameters governing planet formation and evolution.

\section{Discussion}

{ \Will
In addition to the observational and theoretical results
presented above, we conducted a comprehensive series of
tests to verify the robustness of the period--C/O
correlation against potential confounding factors,
including survey heterogeneity, detection biases, and
stellar properties. These tests are detailed in the
following subsections. Although our sample is
heterogeneous, encompassing different facilities used to
measure the C/O ratio and various planetary discovery
methods, the period--C/O trend remains evident (albeit
with reduced statistical significance) across different
subsamples defined by specific facilities and planetary
discovery methods. Furthermore, the period--C/O
correlation is not influenced by other stellar parameters,
including metallicity ($\rm{[Fe/H]}$), effective
temperature ($T_{\text{eff}}$), and surface gravity
($\log g$).}

\subsection{Effect of Heterogeneity in Stellar C/O Measurements}
\label{subsec:survey_heterogeneity}

To assess potential systematic biases arising from the heterogeneous measurement methodologies of the three independent surveys (SDSS-IV, HARPS, and Keck), we repeated our main analysis for each subsample separately.

Regarding the C/O ratio distribution, all three independent subsamples consistently show that cold giant-planet hosts possess higher median C/O ratios compared to hot giant-planet hosts. The strength of this dichotomy, however, differs among surveys: Kolmogorov--Smirnov tests yield $p$-values of $6\times10^{-4}$ ($\sim3.4\sigma$), $2\times10^{-2}$ ($\sim2.3\sigma$), and $0.2$ ($\sim1.4\sigma$) for the SDSS-IV, HARPS, and Keck subsamples, respectively. Thus, the SDSS-IV and HARPS subsamples exhibit a statistically significant C/O dichotomy, while the Keck subsample shows the same trend but with lower significance, likely due to its smaller sample size. Furthermore, the absence of a C/O dichotomy in small-planet hosts ($R_{\rm p} < 4 R_\oplus$) is consistently observed across all individual subsamples, aligning with the results from the combined dataset.

In terms of the period-C/O correlation for giant planets, the SDSS-IV and Keck subsamples exhibit regression slopes of 0.0253 and 0.0202, respectively, which are consistent with the slope derived from the combined sample's 0.0225 in the main text. The HARPS subsample shows a weaker correlation with a slope of 0.0053. This discrepancy is attributed to the selection function of the HARPS GTO program, which contains a significantly lower fraction of short-period giant planets ($32\%$) compared to the SDSS-IV ($75\%$) and Keck ($67\%$) samples, thereby limiting the dynamic range required to constrain the correlation.

Collectively, the consistency of the C/O dichotomy and period trends across these heterogeneous datasets suggests that measurement heterogeneity does not dominantly influence our principal conclusions.

{\Will 
\subsection{Effect of Heterogeneity in Planetary Discovery Method}
\label{sec:dis_method}

Different exoplanet detection methods (Transit vs. Radial Velocity) are sensitive to planets in different orbital regimes. 
As a result, any observed correlation between the host star C/O ratio and orbital period could be biased by differences in period sampling. Here, we perform additional tests to evaluate whether the period-C/O correlation is artificially introduced by the different period distributions of RV and Transit detections.

\begin{figure*}[htbp]
	\centering
	\includegraphics[width=0.9\textwidth]{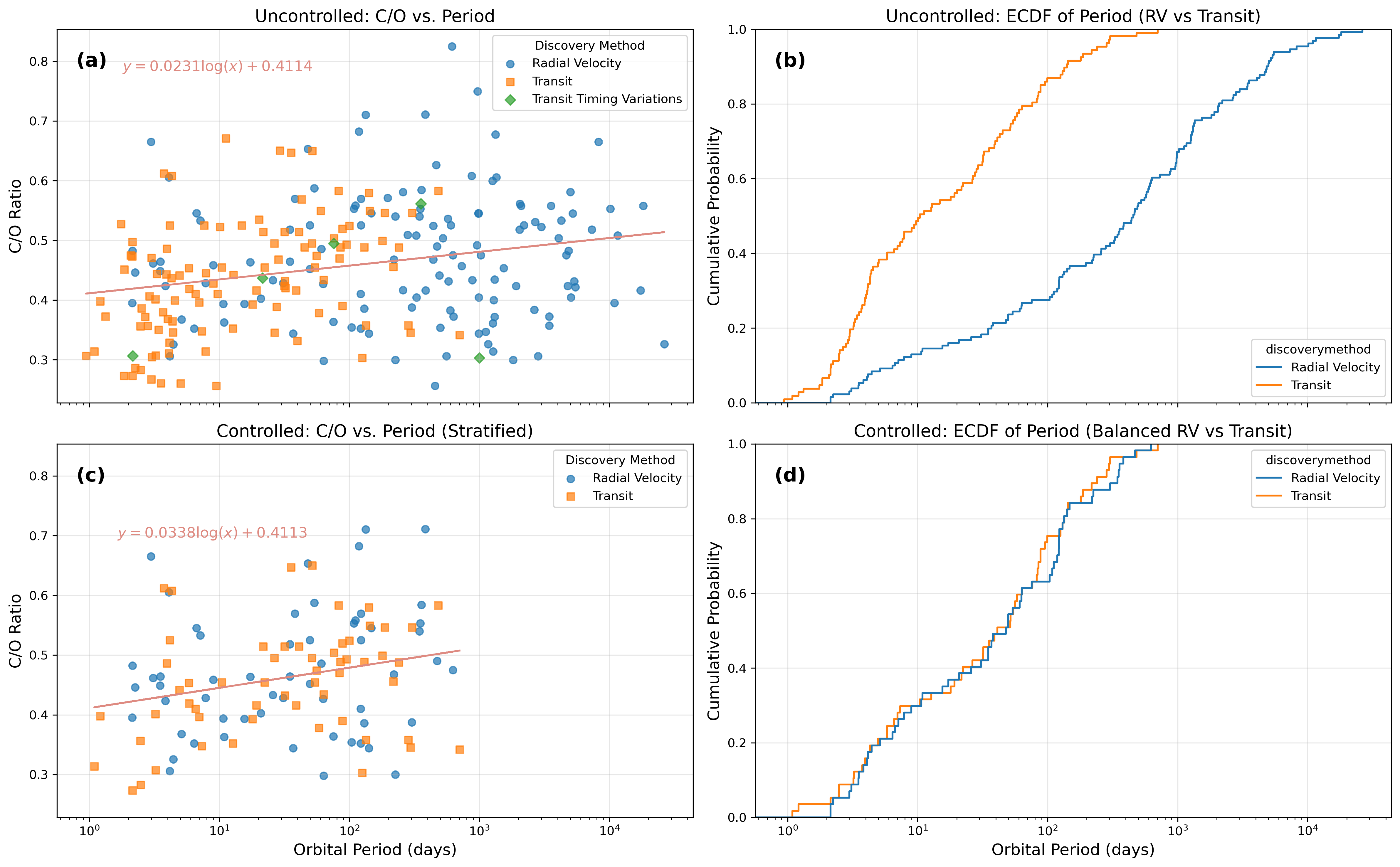} 

	\caption{\textbf{Comparison of C/O--Period Trends in Transit and RV Samples: Uncontrolled vs. Controlled (Stratified).}
        Top row: Uncontrolled sample of giant planets. 
        (a) C/O ratio vs. orbital period for all discovery methods, with a linear regression fit shown in pink. 
        (b) Cumulative distribution functions (CDFs) of orbital periods for Transit and Radial Velocity (RV) planets, highlighting differences in period sampling. 
        Bottom row: Controlled sample with stratified period-matched sub-sampling between RV and Transit samples.
        (c) Regression between C/O ratio and orbital period using the balanced sample.
        (d) Corresponding CDFs confirm similar period distributions in the controlled case. 
        Regression lines are shown with their equations annotated in pink. All regression fits are performed in log-period space.
     }
	\label{fig:Figure_Transit_RV_control} 
\end{figure*}

{In Figure~\ref{fig:Figure_Transit_RV_control}, we compare C/O--period trends before and after controlling for detection-method biases. In the uncontrolled sample of giant planets, panel~(a) shows a modest positive correlation between host-star C/O and $\log_{10} P$ (slope $=0.0231$), while the CDFs in panel~(b) demonstrate that Transit and RV planets occupy different period regimes, with Transit detections skewed toward shorter orbits. To remove this bias, we construct a stratified, period-matched subsample by binning the $\log$-period range and randomly selecting equal numbers of RV and Transit planets within each bin. The resulting balanced sample (panels~c--d) yields matched period distributions, as confirmed by the overlapping CDFs in panel~(d), and the C/O--period regression correspondingly strengthens (slope $=0.0338$; panel~c). Together, these results show that although period-sampling differences exist between detection methods, the positive C/O--period correlation persists—and becomes slightly stronger—once these differences are controlled for.}

To further verify that the correlation is not an artifact of combining datasets, we independently assessed the period--C/O trend within each detection method. We restricted both the Transit and RV giant-planet samples to their overlapping orbital-period range (1--1000 days) to ensure a comparable baseline. 
For the Transit-detected subsample, the difference in C/O between short- and long-period planets ($P < 30$\,d vs.\ $P \ge 30$\,d) is statistically significant ($P$-value = $1\times10^{-2}$, $2.5\sigma$), and the regression slope is 0.0376.
Similarly, the RV sample shows a weaker yet consistent signal ($P$-value = $3\times10^{-2}$, $2.1\sigma$; slope = 0.0331).
Both methods independently show a positive correlation between host star C/O and orbital period. 
This confirms that the observed trend is robust and not an artifact of mixing detection methods.

}

\subsection{Effect of Different Treatments on Multiple-Planet Systems}
\label{sec:Treatments_on_Multiple-Planet_Systems}

In multi-planet systems, several planets with different orbital periods share the same host star and thus the same stellar C/O value. Including multiple planets per system could potentially bias the observed correlation by over-weighting specific stars.
To address this, we adopted a strict selection strategy where only the closest-in planet (typically the most detectable) from each multi-planet system was selected for analysis, ensuring that each host star contributes exactly one data point.

We repeated our main analysis on this restricted sample and found results consistent with the full dataset.
Specifically, a statistically significant difference in C/O is observed between short- and long-period giant planet hosts, yielding a \textit{P}-value of $4 \times 10^{-3}$ ($2.9\sigma$). In contrast, no significant difference is found for small planet hosts ($P = 0.17$), mirroring the behavior of the sample in the main text.
Furthermore, the linear regression recovers a positive slope of $0.0220^{+0.0029}_{-0.0031}$ and a Pearson correlation coefficient of 0.27. Permutation tests indicate that these values are statistically significant, appearing in less than 0.02\% of random trials.

These results confirm that the period--C/O correlation among giant planet hosts persists even after rigorously accounting for planet multiplicity, indicating that the trend is not an artifact of repeated stellar C/O values in multi-planet systems.

\subsection{Effect of Stellar Properties}

To ensure that the observed period--C/O correlation is not driven by underlying stellar properties, we investigated the potential confounding effects of stellar metallicity ([Fe/H]), effective temperature ($T_{\rm eff}$), and surface gravity ($\log g$). We employed a stratified binning with matched sub-sampling technique \citep{cochran1977sampling} to construct controlled subsamples where the distributions of the target stellar property were statistically indistinguishable between hot ($P < 30$\,d) and warm ($P > 30$\,d) giant planet hosts.

{\Will 
\subsubsection{Stellar Metallicity}
Stellar metallicity is known to correlate with planetary occurrence and elemental ratios \citep{2018A&A...614A..84S, 2021A&A...655A..99D}. In our uncontrolled sample, the [Fe/H] distributions of hot and warm giant planet hosts showed a moderate difference ($P = 0.03$), while the C/O distributions exhibited a significant offset ($P = 5.7 \times 10^{-6}$). These distributions are illustrated in the left panels of Figure~\ref{fig:feh_co}, where the cumulative distribution functions (CDFs) of [Fe/H] for hot (orange) and warm (green) giant planet hosts show a visible separation, and the scatter plot of C/O versus [Fe/H] reveals that warm giant hosts are systematically shifted toward higher C/O values across the full metallicity range.

After applying the stratified matching procedure, the [Fe/H] difference was effectively eliminated ($P = 0.77$), as shown by the nearly overlapping CDFs in the upper-right panel of Figure~\ref{fig:feh_co}. Crucially, a highly significant difference in C/O ratios persisted in this metallicity-controlled sample ($P = 2.5 \times 10^{-6}$; lower-right panels of Figure~\ref{fig:feh_co}). The scatter plot confirms that the vertical separation in C/O between the two populations is preserved even after the [Fe/H] distributions have been matched. This demonstrates that the observed C/O trend is independent of host star metallicity and is not a secondary consequence of the well-known planet--metallicity correlation.

\begin{figure*}[ht!]
\centering
\includegraphics[width=\textwidth]{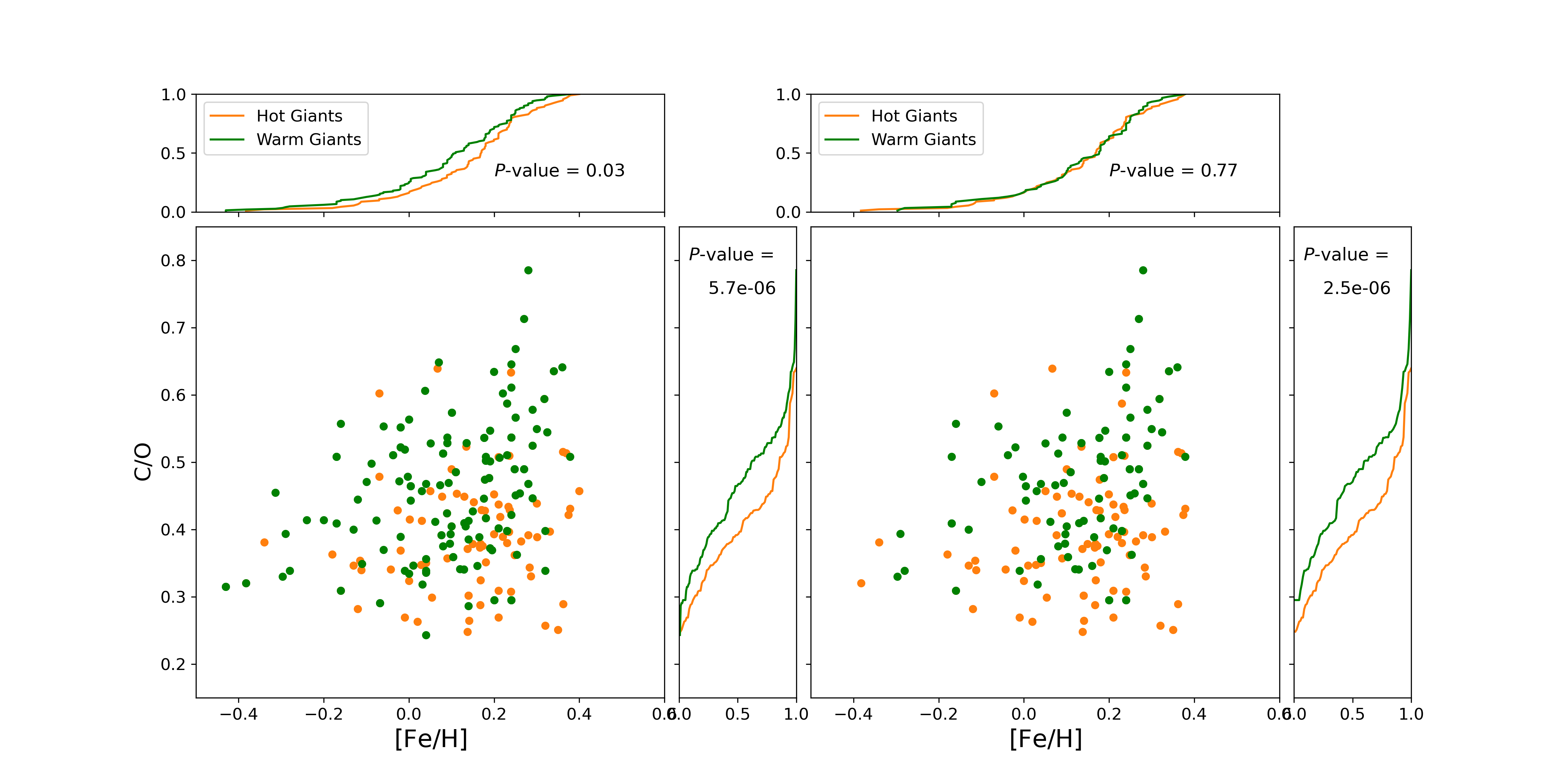}
\caption{Effect of stellar metallicity control on the C/O dichotomy between hot and warm giant planet hosts. \textbf{Left panels:} Uncontrolled sample. The upper-left panel shows the CDFs of [Fe/H] for hot (orange) and warm (green) giant planet hosts, with a moderate difference ($P = 0.03$). The lower-left panel displays C/O versus [Fe/H], with the adjacent CDF panel showing a significant C/O offset between the two populations ($P = 5.7 \times 10^{-6}$). \textbf{Right panels:} Metallicity-controlled sample after stratified matching. The upper-right panel confirms that the [Fe/H] distributions are now statistically indistinguishable ($P = 0.77$). The lower-right panel shows that the C/O difference remains highly significant ($P = 2.5 \times 10^{-6}$), demonstrating that the C/O dichotomy is independent of stellar metallicity.}
\label{fig:feh_co}
\end{figure*}
}

\subsubsection{Effective Temperature}
Regarding stellar effective temperature, the uncontrolled sample already exhibited broadly similar $T_{\rm eff}$ distributions between hot and warm hosts ($P = 0.45$). Nevertheless, to rigorously rule out any temperature bias, we applied the control procedure, yielding matched $T_{\rm eff}$ distributions ($P = 0.94$). In this controlled sample, the difference in C/O ratios remained highly significant with a $P$-value of $6.5 \times 10^{-4}$ (corresponding to $3.4\sigma$). These results demonstrate that the C/O enhancement in warm giant hosts is not driven by systematic differences in $T_{\rm eff}$.

\subsubsection{Surface Gravity}
Finally, we investigated surface gravity ($\log g$), which can influence inferred chemical abundances through atmospheric effects. The initial sample showed a significant difference in $\log g$ between hot and warm hosts ($P = 0.02$). However, after constructing a controlled sample where the $\log g$ distributions were statistically identical ($P = 0.66$), the C/O distributions remained statistically distinct ($P = 5.6 \times 10^{-5}$, corresponding to $4.0\sigma$).
This reinforces the conclusion that the higher C/O ratios observed in warm giant hosts are not an artifact of differing stellar surface gravities.

In summary, the persistence of the period--C/O correlation across these controlled subsamples provides strong evidence that the trend is not a secondary manifestation of stellar atmospheric parameters ([Fe/H], $T_{\rm eff}$, or $\log g$).

\subsection{Effect of C/O Uncertainty}

To evaluate the robustness of our results against measurement errors, we conducted a Monte Carlo bootstrap analysis. 
We generated 100,000 synthetic catalogs by resampling the [C/H] and [O/H] abundances of each star from normal distributions defined by their reported values and associated measurement uncertainties. 
For each realization, we performed a two-sample Kolmogorov–Smirnov (K–S) test to compare the C/O distributions of hot ($P < 30$\,d) and warm ($P > 30$\,d) giant planet hosts.

The resulting distribution of $P$-values yields a median of $5.0 \times 10^{-6}$, with a 68.3\% confidence interval spanning from $1.4 \times 10^{-6}$ to $2.4 \times 10^{-5}$. 
This demonstrates that the statistical distinction between the two populations remains robust at the $\sim4\sigma$ level even when fully accounting for measurement uncertainties.

We note that the median bootstrap $P$-value is slightly more conservative than the deterministic result derived from the nominal values. 
This is expected because the resampling process introduces random noise that effectively broadens the underlying C/O distributions of both populations. 
This broadening naturally increases the distributional overlap, thereby reducing the maximum distance between cumulative distribution functions (the K–S statistic $D$) and systematically elevating the resulting $P$-values. 
Consequently, this analysis confirms that the observed C/O dichotomy is not an artifact of measurement noise.

\subsection{Effects of Abundance Calibration on the C/O–Period Relation}
\label{subsec:abundance_calibration}

To assess how inter-survey calibration affects the derived C/O–period trends, we performed a systematic calibration of stellar carbon and oxygen abundances among the three major spectroscopic datasets used in this work—\textsc{SDSS}, \textsc{Keck}, and \textsc{HARPS}.
We adopted the ordinary least-squares bisector (OLS-bisector) regression method \citep{Isobe1990}, which treats both variables symmetrically and is robust against measurement errors in both axes.

{\Will 
Before applying any calibration, we first examined the raw agreement between surveys using stars observed in common. Our overlap samples consist of 9 stars for SDSS--HARPS, 69 stars for SDSS--Keck, and 80 stars for HARPS--Keck. The HARPS--Keck and SDSS--Keck pairs show small systematic offsets (mean deviations $\lesssim 0.03$~dex) and modest scatter (standard deviations of 0.06--0.13~dex) in both [C/H] and [O/H]. The HARPS--SDSS comparison is similarly consistent for [C/H] (MD $= 0.02$~dex), but reveals a larger offset in [O/H] (MD $= 0.14$~dex, Std $= 0.21$~dex; Figure~\ref{fig:delta_ch_oh_HARPS_sdss}), attributable to the well-known systematics between optical O\,\textsc{i} triplet and near-infrared OH-based oxygen measurements. The limited overlap between SDSS and HARPS (9 stars) also contributes to the larger scatter in this pair.}

\subsubsection{Calibration to SDSS-IV}
We derived empirical calibration relations between \textsc{Keck} and \textsc{SDSS-IV} using common stars observed by both surveys (Figure~\ref{fig:delta_ch_oh_SDSS_vs_Keck_calibrated}). The resulting relations are:
\begin{equation}
\begin{aligned}
[\mathrm{C/H}]_{\mathrm{SDSS}} &= 1.24\,[\mathrm{C/H}]_{\mathrm{KECK}} - 0.04, \\
[\mathrm{O/H}]_{\mathrm{SDSS}} &= 1.39\,[\mathrm{O/H}]_{\mathrm{KECK}} - 0.005 .
\end{aligned}
\end{equation}
Applying these transformations slightly steepens the C/O–period slope from $0.0225$ to $0.0242$ and enhances the statistical significance of the hot–cold giant planet separation (KS $P$-value decreases from $6\times10^{-6}$ to $5\times10^{-6}$). This enhancement is a natural consequence of the calibration slopes being greater than unity, which broaden the dynamic range of the abundances.
For \textsc{HARPS}, due to the small overlap with \textsc{SDSS} (9 stars; Figure~\ref{fig:delta_ch_oh_HARPS_sdss}), no explicit calibration was applied to avoid overfitting.

\subsubsection{Calibration to Keck}
{\Will As noted above, combining NIR and optical abundances introduces methodological challenges that motivate explicit cross-calibration. }To robustly account for these intrinsic methodological differences and potential zero-point offsets, we calibrated \textsc{HARPS} and \textsc{SDSS} abundances onto the \textsc{Keck} scale using an Ordinary Least-Squares (OLS) bisector regression on a mutually observed stellar subsample (Figures~\ref{fig:delta_ch_oh_HARPS_vs_Keck_calibrated} and \ref{fig:delta_ch_oh_Keck_vs_SDSS_calibrated}). The derived empirical relations for \textsc{HARPS}-to-\textsc{Keck} are:
\begin{equation}
\begin{aligned}
[\mathrm{C/H}]_{\mathrm{KECK}} &= 0.92\,[\mathrm{C/H}]_{\mathrm{HARPS}} + 0.01, \\
[\mathrm{O/H}]_{\mathrm{KECK}} &= 0.88\,[\mathrm{O/H}]_{\mathrm{HARPS}} + 0.04 .
\end{aligned}
\end{equation}
And for \textsc{SDSS}-to-\textsc{Keck}:
\begin{equation}
\begin{aligned}
[\mathrm{C/H}]_{\mathrm{KECK}} &= 0.81\,[\mathrm{C/H}]_{\mathrm{SDSS}} + 0.04, \\
[\mathrm{O/H}]_{\mathrm{KECK}} &= 0.72\,[\mathrm{O/H}]_{\mathrm{SDSS}} + 0.032 .
\end{aligned}
\end{equation}

The slopes are relatively close to unity for both carbon and oxygen, confirming that the three surveys are generally consistent in their abundance scales. Crucially, the empirical calibration accounts for the systematic zero-point offsets, which are remarkably small (of order $\sim0.01$–$0.04$ dex).

Because some of these slopes (particularly for SDSS-to-Keck) are less than unity, transforming to the \textsc{Keck} scale mathematically compresses the dynamic range of the abundances. Consequently, the C/O–period slope decreases to $0.0113$, and the significance of the hot–cold giant separation softens (the K-S $P$-value increases to $3\times10^{-3}$). However, the correlation remains statistically significant ($>2\sigma$), indicating that while the compression affects the amplitude of the signal, the qualitative presence of the trend is highly robust against inter-survey systematic differences.

\subsubsection{Calibration to HARPS}
Finally, calibrating \textsc{Keck} abundances onto the \textsc{HARPS} system (Figure~\ref{fig:delta_ch_oh_Keck_vs_HARPS_calibrated}) yields relations with slopes close to unity:
\begin{equation}
\begin{aligned}
[\mathrm{C/H}]_{\mathrm{HARPS}} &= 1.09\,[\mathrm{C/H}]_{\mathrm{KECK}} - 0.01, \\
[\mathrm{O/H}]_{\mathrm{HARPS}} &= 1.14\,[\mathrm{O/H}]_{\mathrm{KECK}} - 0.04 .
\end{aligned}
\end{equation}
Because the transformation is nearly linear with unit slope, the resulting C/O distribution and period trend remain virtually unchanged (slope shifts marginally to $0.0222$; KS $P$-value shifts to $3\times10^{-6}$). Direct calibration between \textsc{SDSS} and \textsc{HARPS} was again omitted due to insufficient overlap.

In summary, across all calibration schemes, the positive correlation between stellar C/O ratio and planetary orbital period persists. {\Will Depending on the calibration reference frame adopted, the K-S $P$-value ranges from $3\times10^{-6}$ (HARPS scale) to $3\times10^{-3}$ (Keck scale), corresponding to significance levels of $\sim$2.7--4.7$\sigma$.} While the absolute strength and significance of the trend scale with the dynamic range of the adopted abundance system, the fundamental astrophysical signal remains robust against inter-survey calibration choices.

{\Will 
To further assess the impact of poorly cross-calibrated 
stars, we repeated our analysis after removing outliers 
with absolute residuals exceeding $2\sigma$ from the 
OLS-bisector fits. For the SDSS-to-Keck calibration, 
this excludes 3 stars in [C/H] and 6 in [O/H]. After 
removal, the median C/O values for the hot and cold 
giant populations shift by less than 0.01~dex, and the 
K--S $P$-value decreases from $6\times10^{-6}$ to 
$2\times10^{-6}$. The log-linear slope shifts only 
marginally from 0.0242 to 0.0245. Similarly, applying 
$2\sigma$ outlier rejection to the Keck-scale calibration 
yields a K--S $P$-value of $9\times10^{-3}$ 
($\sim2.6\sigma$), with the slope shifting from 0.0113 
to 0.0104. These tests confirm that the poorly 
cross-calibrated values do not drive our conclusions.
}

\begin{figure*}[htbp]
    \centering
    
    \begin{minipage}{0.48\textwidth}
        \centering
        \includegraphics[width=\linewidth]{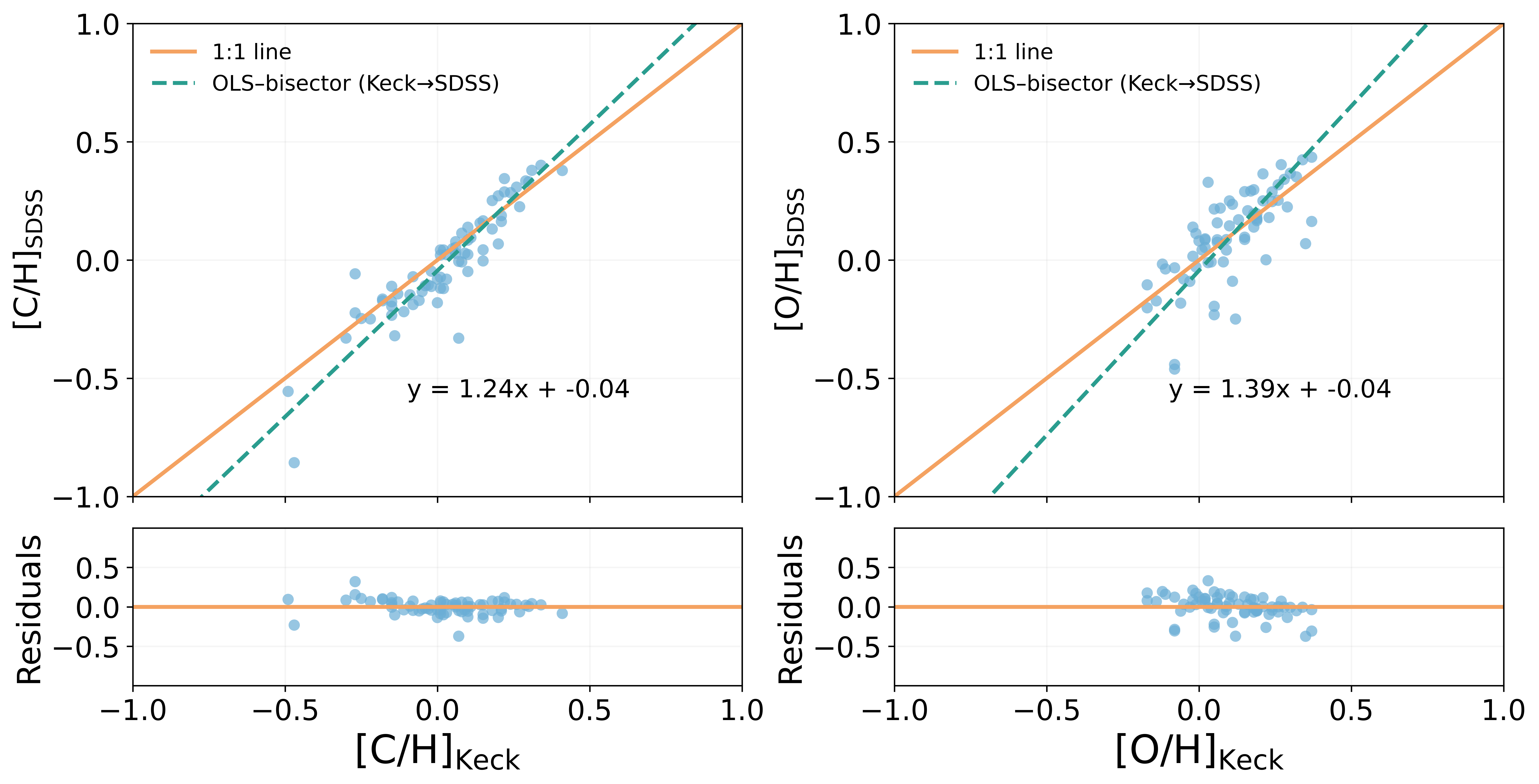}
        \caption{
            \textbf{Cross-calibration of [C/H] and [O/H] between the \textsc{Keck} and \textsc{SDSS} samples.}
            \textbf{Top:} Comparison of abundances between the two surveys. The orange solid line indicates the 1:1 relation, while the teal dashed line represents the best-fit OLS-bisector calibration.
            \textbf{Bottom:} Residuals relative to the best-fit line.
        }
        \label{fig:delta_ch_oh_SDSS_vs_Keck_calibrated}
    \end{minipage}\hfill
    \begin{minipage}{0.48\textwidth}
        \centering
        \includegraphics[width=\linewidth]{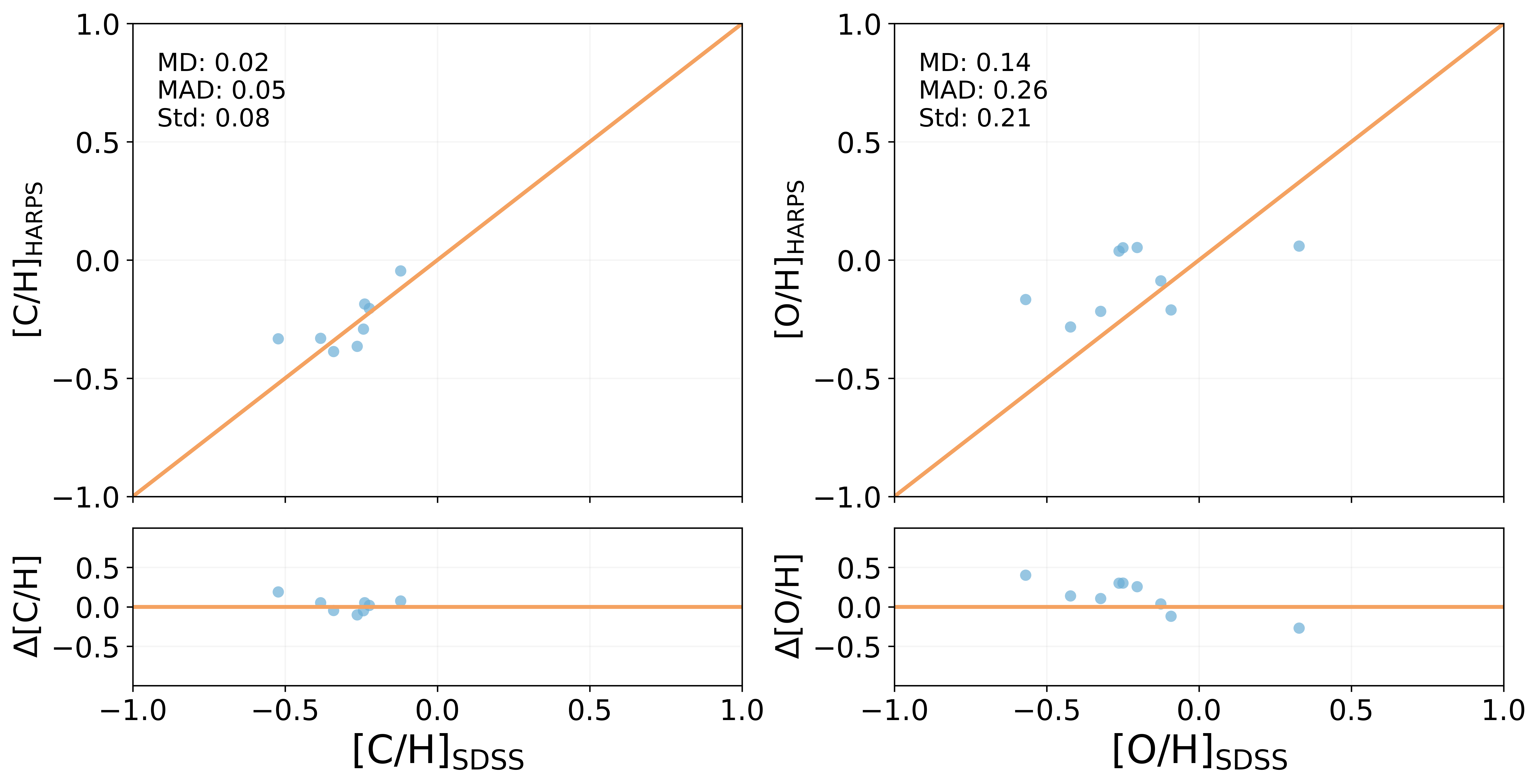}
        \caption{
            \textbf{Comparison of stellar abundances between \textsc{HARPS} and \textsc{SDSS}.}
            \textbf{Top:} Direct comparison of [C/H] (left) and [O/H] (right). The orange line indicates the 1:1 relation.
            \textbf{Bottom:} Residuals relative to the 1:1 line.
            Inset statistics denote the mean deviation (MD), median absolute deviation (MAD), and standard deviation (Std).
        }
        \label{fig:delta_ch_oh_HARPS_sdss}
    \end{minipage}
    
    \vspace{0.8cm} 
    
    \begin{minipage}{0.48\textwidth}
        \centering
        \includegraphics[width=\linewidth]{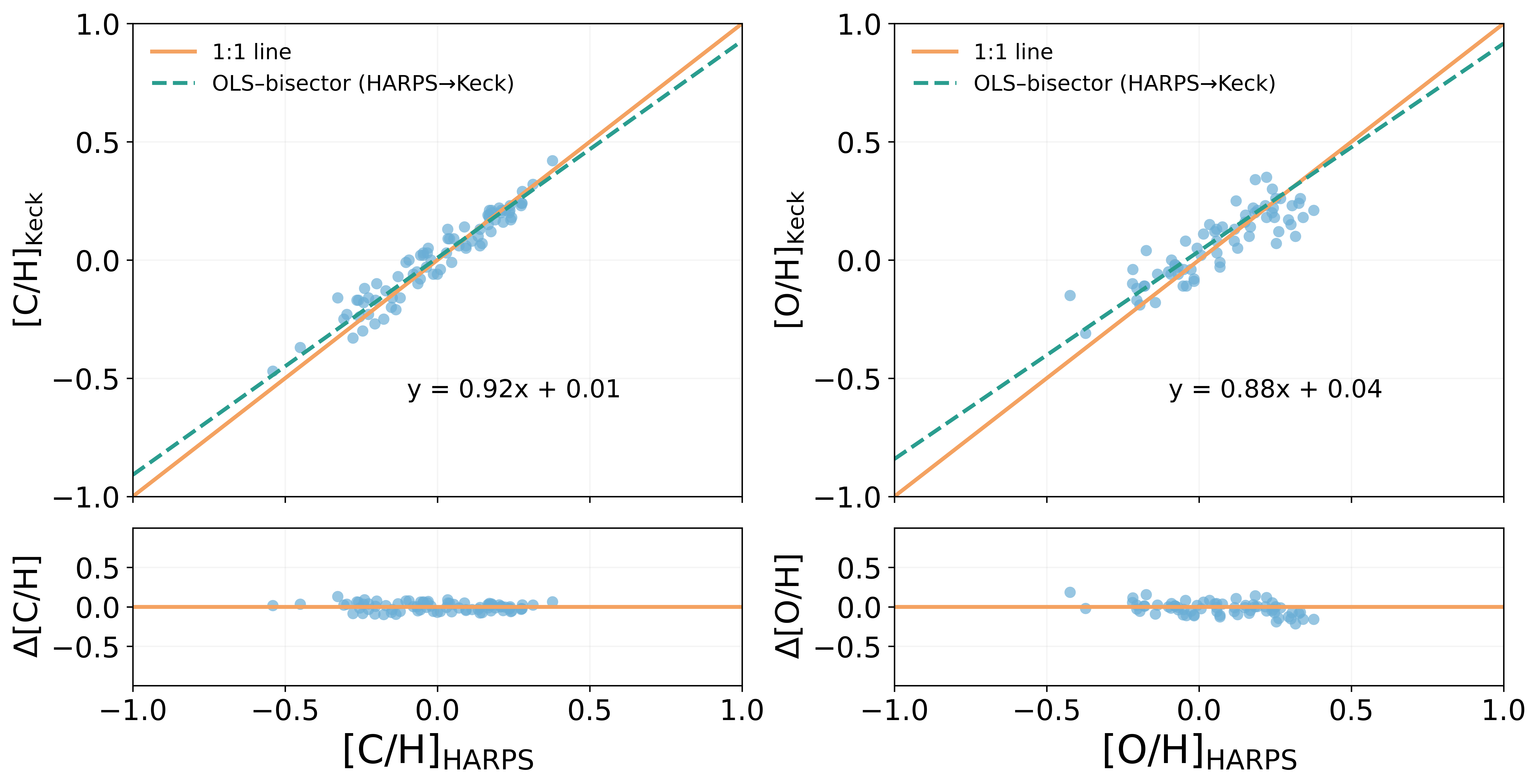}
        \caption{
            \textbf{Calibration of abundances from \textsc{HARPS} to \textsc{Keck}.}
            \textbf{Top:} Correlations of [C/H] (left) and [O/H] (right). The orange solid line indicates the 1:1 relation, and the teal dashed line shows the best-fit OLS-bisector calibration.
            \textbf{Bottom:} Residuals relative to the best-fit line.
        }
        \label{fig:delta_ch_oh_HARPS_vs_Keck_calibrated}
    \end{minipage}\hfill
    \begin{minipage}{0.48\textwidth}
        \centering
        \includegraphics[width=\linewidth]{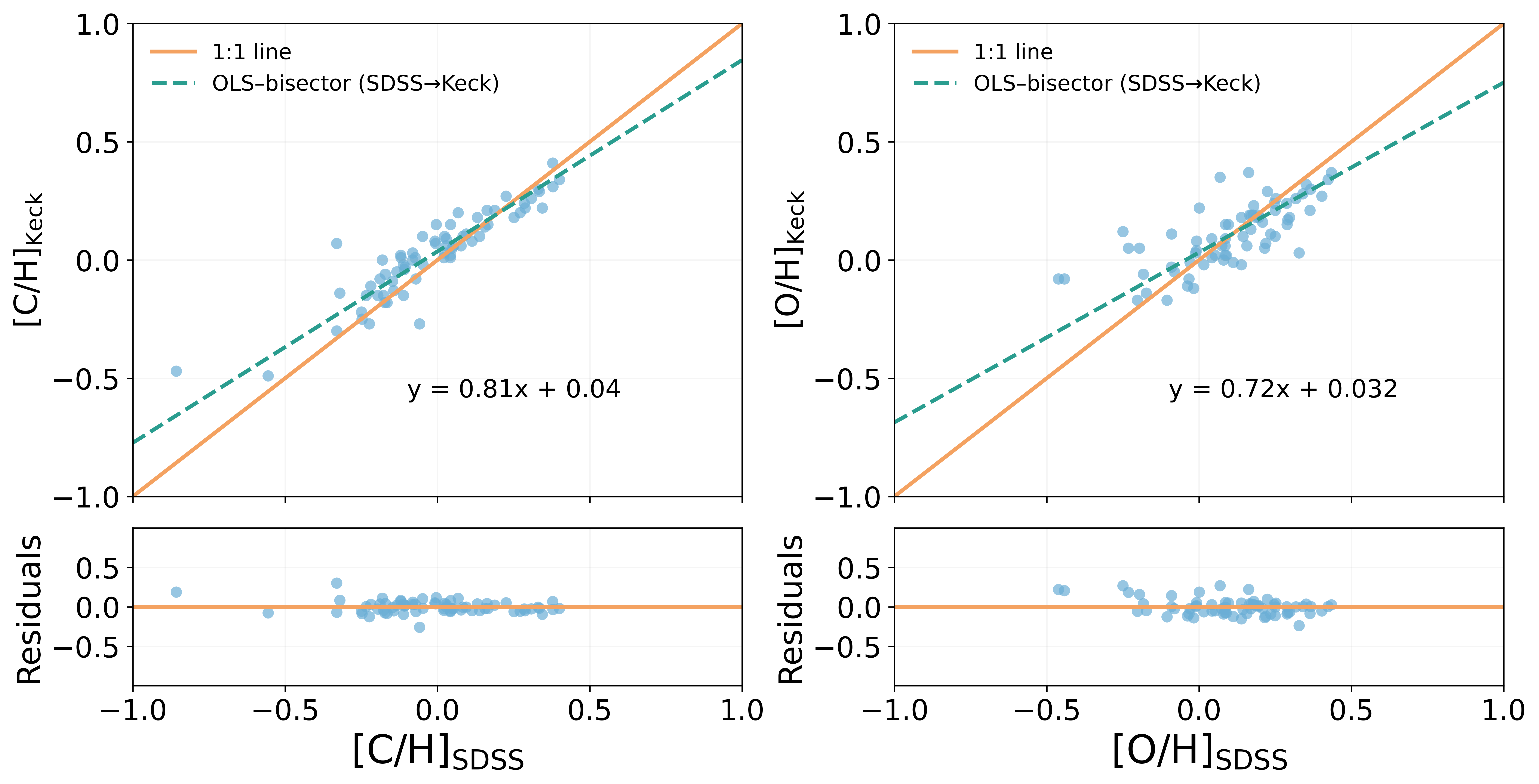}
        \caption{
            \textbf{Calibration of abundances from \textsc{SDSS} to \textsc{Keck}.}
            \textbf{Top:} Correlations of [C/H] (left) and [O/H] (right). The orange solid line indicates the 1:1 relation, and the teal dashed line shows the best-fit OLS-bisector calibration.
            \textbf{Bottom:} Residuals relative to the best-fit line.
        }
        \label{fig:delta_ch_oh_Keck_vs_SDSS_calibrated}
    \end{minipage}

    \vspace{0.8cm} 

    \begin{minipage}{0.48\textwidth}
        \centering
        \includegraphics[width=\linewidth]{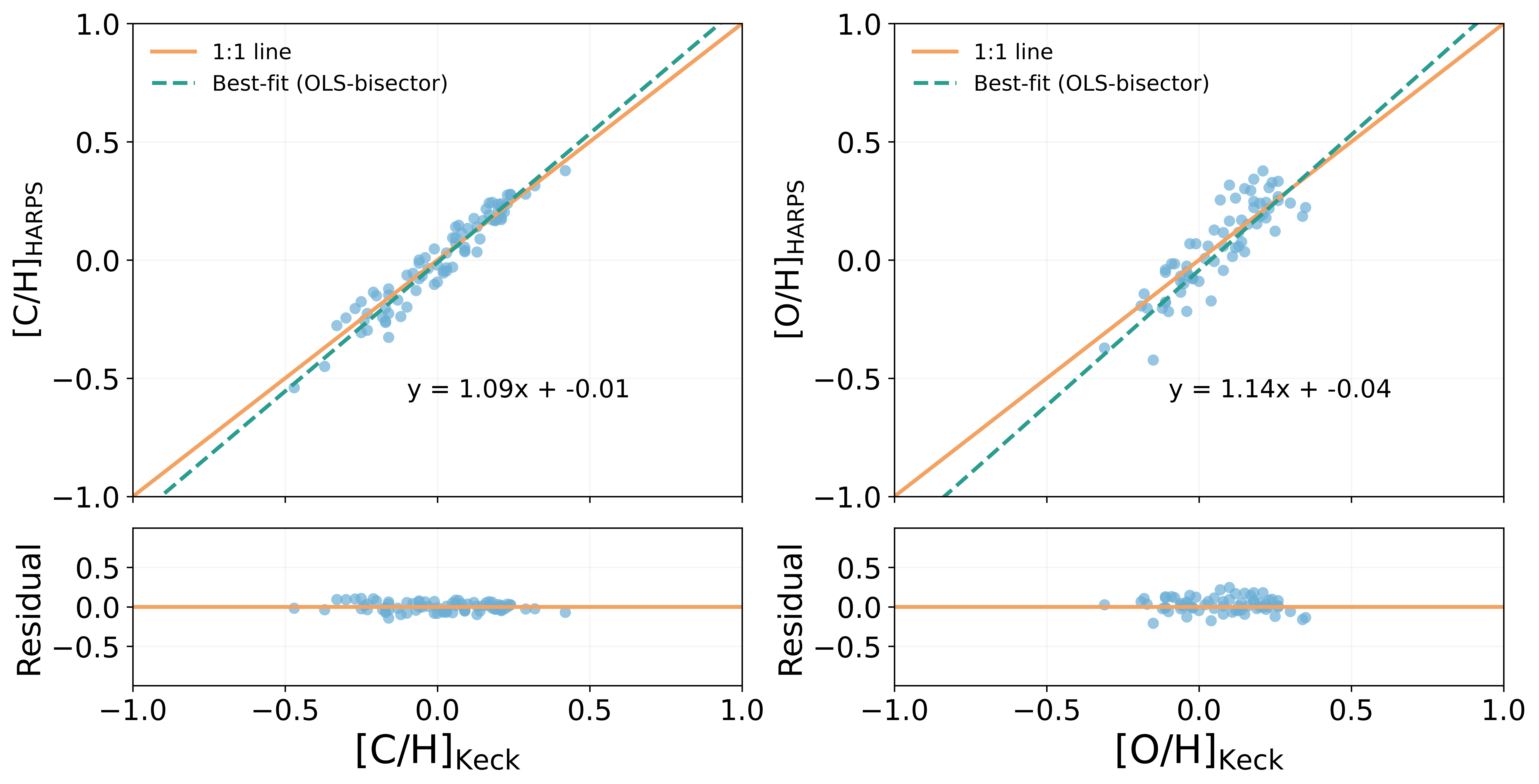}
        \caption{
            \textbf{Calibration of abundances from \textsc{Keck} to \textsc{HARPS}.}
            \textbf{Top:} Correlations of [C/H] (left) and [O/H] (right). The orange solid line indicates the 1:1 relation, and the teal dashed line shows the best-fit OLS-bisector calibration.
            \textbf{Bottom:} Residuals relative to the best-fit line.
        }
        \label{fig:delta_ch_oh_Keck_vs_HARPS_calibrated}
    \end{minipage}
    
\end{figure*}

{\Will

\subsection{Caveats and future studies}

We note that our model adopts several simplifications. First, the 
quantitative fraction of short-period giants is slightly lower than 
observed; reproducing the full Hot Jupiter pile-up would require 
additional mechanisms such as magnetospheric truncation or post-disk 
High-Eccentricity tidal Migration (HEM) driven by planet--planet 
scattering or Kozai--Lidov resonances \citep{2018ARA&A..56..175D}. 
Such pathways can transport high-C/O planets from the outer disk to 
short-period orbits, contributing to population mixing, though the low 
probability of these channels---consistent with the observed rarity of 
Hot Jupiters \citep[$\sim$0.5--1.2\%;][]{Wright2012, Beleznay2022}---ensures 
that the C/O--period correlation remains statistically robust. Second, 
our treatment of ice line physics is simplified: in reality, ice lines 
migrate inward as the disk cools \citep[e.g.,][]{Garaud2007} and 
represent two-dimensional surfaces in the ($r$, $z$) plane. However, 
since dust settles rapidly toward the midplane \citep{Dullemond2004, 
Villenave2020} and the relative spatial ordering of ice lines is 
conserved---H$_2$O always condenses interior to CO$_2$ and organic 
carbon---the key mechanism driving the C/O--period correlation depends 
on this relative gradient rather than on precise radial locations. 
The dynamic nature of ice lines would naturally introduce additional 
scatter into the final orbital periods, consistent with the observed 
spread (Figure~\ref{fig:Figure3}). Incorporating time-dependent ice 
line evolution and HEM channels would refine our quantitative 
predictions, and we defer such treatments to future work.
}

{\Will
\section{CONCLUSION}
\label{sec:conclusion}

In this study, we investigated the influence of the host star carbon-to-oxygen (C/O) ratio on the orbital architectures of planetary systems. By compiling a unified and cross-calibrated catalog of 598 planet-hosting stars (harboring 929 planets) from the SDSS-IV/APOGEE, Keck, and HARPS surveys, we explored the statistical relationship between stellar volatile abundances and planetary orbital periods. Our primary findings and interpretations are summarized as follows:

\renewcommand{\labelenumi}{(\arabic{enumi})}
\begin{enumerate}
    \item \textbf{A C/O--Period Correlation for Giant Planets:} We identified a statistically significant positive correlation between the host star C/O ratio and the orbital periods of giant planets ($R_P \ge 4\,R_\oplus$). Giant planets residing in longer-period orbits ($P \ge 30$\,days) systematically orbit stars with higher C/O ratios (median $\sim0.47$) compared to their short-period hot Jupiter counterparts (median $\sim0.39$). Extensive control tests confirm that this C/O dichotomy is not an artifact of observational biases, heterogeneous detection methods (Transit vs.\ Radial Velocity), or underlying stellar properties (such as [Fe/H], $T_{\rm eff}$, and $\log g$). Furthermore, our two-dimensional parameter space exploration demonstrates that this correlation occupies a broad, stable region of high statistical significance -- particularly along the period axis—consistent with a continuous C/O--period trend rather than a fragile, fine-tuned threshold.
    
    \item \textbf{A Distinct Formation Pathway from Small Planets:} In stark contrast to the giant planet population, small planets ($R_P < 4\,R_\oplus$) exhibit similar host-star C/O ratios regardless of their orbital periods. This divergence suggests that the stellar C/O ratio primarily governs the formation and migration pathways of massive gas giants, leaving little imprint on the spatial distribution of inner, smaller planets.
    
    \item \textbf{A Pebble-Driven Theoretical Framework:} We utilized a pebble-driven core accretion population synthesis model to provide a physical interpretation for this observed trend. We demonstrate that the stellar C/O ratio—acting as a proxy for the protoplanetary disk's volatile partitioning—fundamentally alters the spatial distribution of solid materials. In high-C/O environments, oxygen is depleted, pushing the dominant solid mass reservoirs outward to the carbon-rich ice lines (e.g., CO$_2$ and organics). Consequently, giant planets initiate formation further out in the disk, undergo moderate inward migration, and ultimately stall at longer orbital periods. Conversely, low-C/O environments concentrate solid mass near the inner water ice line, facilitating the formation and subsequent migration of close-in hot giant planets.
\end{enumerate}

Finally, our results establish the stellar C/O ratio as a critical astrophysical lever that shapes the macroscopic orbital architecture of giant planetary systems. The chemical partitioning of the natal molecular cloud leaves an indelible, predictable imprint on the final planetary configuration. Future high-precision spectroscopic surveys of exoplanet hosts, coupled with advanced hydrodynamical simulations incorporating dynamic ice lines and high-eccentricity migration, will be essential to further unravel the complex, multidimensional interplay between disk chemistry and planet formation.
}


\begin{acknowledgments}
This work is supported by the National Key R\&D Program of China (No. 2024YFA1611803) and the National Natural Science Foundation of China (NSFC; Grant Nos. 12273011, 12150009, 12403071). J.-W.X. also acknowledges the support from the National Youth Talent Support Program. 

This research has used the NASA Exoplanet Archive, which is operated by the California Institute of Technology, under contract with the National Aeronautics and Space Administration under the Exoplanet Exploration Program \citep{2025arXiv250603299C}.
\end{acknowledgments}

\appendix

\renewcommand{\thetable}{A\arabic{table}}
\setcounter{table}{0}

\section{Simulations of Population Synthesis Model}
\label{sec:pop_synthesis} 

We conduct a pebble-driven core accretion planet population synthesis model to investigate the formation and evolution of giant planets around stars with varying C/O abundance \citep{Liu2019}. The key physical processes and equations are summarized as follows, whereas other model details can be found in Section 2 of \cite{Liu2019}. 

The planet growth includes the accretion of solids and gas. The growth of the core proceeds through pebble accretion, assuming that dust particles in disks have rapidly grown to one-millimeter-sized pebbles. The corresponding mass growth rate can be expressed as

\begin{equation}
\begin{split}  
  \dot{M}_{\rm PA} &= \varepsilon_{\rm PA} \dot{M}_{\rm peb} = \varepsilon_{\rm PA} \xi_{\rm p/g}  \dot{M}_{\rm g} \\
  &= \left(\varepsilon_{\rm PA,2D}^{-2} + \varepsilon_{\rm PA,3D}^{-2}\right)^{-1/2}  \xi_{\rm p/g}  \dot{M}_{\rm g},
\end{split}
\end{equation}

where $\dot{M}_{\rm peb}$ and $\dot M_{\rm g}$ are the pebble and gas mass flux,  and $\varepsilon_{\rm PA}$ is the total pebble accretion efficiency. The pebble and gas flux ratio can be expressed as 
\begin{equation}
 \xi_{\rm p/g} = \dot M_{\rm peb} / \dot M_{\rm g} = Z_{\odot} \times 10^{\rm Fe/H} f_{\rm cdf, ice},
\end{equation}
which is assumed to be a fixed value at the outermost disk region, and decrease as pebbles drift across and sublimate at different volatile ice lines, $Z_{\odot}=0.014$ and $f_{\rm cdf, ice}$ is the cumulative mass fraction of different chemical species for a solar metallicity disk (calculated from the third column of Table  \ref{tab:chemistry}). 

Following the disk chemical composition model for Sun-like stars adopted in \cite{Turrini2021}, we assume that $ 48\%$ of total O is locked up into refractory rocks, $32\%$ of O is in $\rm H_2O$, $ 13.5\%$ of O is in $\rm CO_2$, and $ 6.5\%$ of O is locked in CO. On the other hand, $9\%$ of C is in refractory rocks, $ 64\%$ is in organic carbon, $ 11.5\%$ is in $\rm CO_2$, $ 4\%$ is in $\rm CH_4$, and $ 11.5\%$ is in CO.  For stars with a different C/O ratio, we assume that the total C and O abundance remain fixed. 
We prioritize the allocation of C or O to $\rm CO_2$, $\rm CH_4$, CO and refractory rocks, while the remaining C and O are assumed to reside in organic carbon and $\rm H_2O$, respectively. The mass fractions of all chemical species are listed in Table \ref{tab:chemistry}. We note that a higher stellar C/O ratio means less solid near the water ice line. 

The 2D and 3D accretion efficiencies are given by \cite{Liu2018} and \cite{Ormel2018}:
\begin{equation}
\begin{split}
\varepsilon_{\rm PA,2D} &= \frac{0.32}{\eta} \sqrt{ \frac{m}{M_{\star}} \frac{1}{\tau_{\rm s}} \frac{\Delta v}{v_{\rm K}}}, \\  
\varepsilon_{\rm PA,3D} &= \frac{0.39}{\eta h_{\rm peb}} \frac{m}{ M_{\star}},
\end{split}
\end{equation}
where $m$ is the planet mass, $M_{\star}$ is the stellar mass, $v_{\rm K}{\equiv}\Omega_{\rm K} r$ is the Keplerian velocity, $\Omega_{\rm K}$ is the angular velocity, $r$ is the radial distance to the star,  $\Delta v$ is the relative velocity between the pebbles and planet,  govern by $\eta v_{\rm K}$ in the headwind regime or $\Omega_{\rm K} R_{\rm H}$ in the shear regime, where $R_{\rm H}{=}(m/3M_{\star})^{1/3}r$ is the planet Hill radius, $\eta = -0.5 h_{\rm g}^2 (\partial \ln P /\partial \ln r)$ is the headwind prefactor, $h_{\rm g}$ is gas disk aspect ratios and $P$ is the gas pressure. The pebble disk aspect ratio is $h_{\rm peb}{ =} \sqrt{\alpha_{\rm t}/(\alpha_{\rm t}+\tau_{\rm s}}) \ h_{\rm g}$, and $\tau_{\rm s}{=}\sqrt{2 \pi}\rho_{\bullet} R_{\rm peb}/\Sigma$ is the pebbles' Stokes number, where $\rho_{\bullet}$, $R_{\rm peb}$ and $\Sigma$ are the internal density, radius of pebbles and gas disk surface density. 

We assume that the disk's evolution is predominately driven by internal viscosity and stellar photoevaporation. The layered  accretion disk we consider follows a two-$\alpha$ scheme, with a quiescent midplane and the upper active, turbulent zone. Specifically, the evolution of $\Sigma$ and $\dot M_{\rm g}$ is coupled with the global disk evolution through the viscous parameter $\alpha_{\rm g}$. On the other hand, $\alpha_{\rm t}$ quantifies the local turbulent diffusion coefficient, roughly corresponding to the turbulent strength near the midplane when the disk turbulence is governed by the magneto-rotational instability (for further clarifications on the two-$\alpha$ approach, refer to \citealt{Liu2019}).

\startlongtable
\begin{deluxetable}{c c l c c c}
\tablecaption{Condensation temperatures and mass fractions of chemical species in our disk model.\label{tab:chemistry}}
\tablewidth{0pt} 
\tablehead{
\colhead{Species} & \colhead{$T_{\rm cond}$ (K)} & \colhead{Mass fraction} &
\colhead{C/O=0} & \colhead{C/O=0.5} & \colhead{C/O=1}
}
\startdata
Silicate & 1500 & $3.5\times10^{-3} + 48\%N_{\rm O}\times16 + 9\%N_{\rm C}\times12$ & $7.6\times10^{-3}$ & $6.7\times10^{-3}$ & $5.9\times10^{-3}$ \\
H$_2$O   & 150  & $16\times(52\%N_{\rm O} - 2N_{\rm CO_2} - N_{\rm CO})$            & $5.0\times10^{-3}$ & $2.4\times10^{-3}$ & $1.8\times10^{-3}$ \\
Org.\ C  & 100  & $12\times(87\%N_{\rm O} - N_{\rm CO_2} - N_{\rm CO})$             & 0                 & $1.5\times10^{-3}$ & $2.7\times10^{-3}$ \\
NH$_3$   & 80   & $3.0\times10^{-4}$                                                & $3.0\times10^{-4}$  & $3.0\times10^{-4}$  & $3.0\times10^{-4}$ \\
CO$_2$   & 65   & $\min(6.5\%N_{\rm O}\times44,\ 11.5\%N_{\rm C}\times44)$           & 0                 & $1.0\times10^{-3}$ & $9.0\times10^{-3}$ \\
CH$_4$   & 30   & $4\%N_{\rm C}\times16$                                             & 0                 & $1.0\times10^{-4}$   & $1.7\times10^{-4}$ \\
CO       & 20   & $\min(6.5\%N_{\rm O}\times28,\ 11.5\%N_{\rm C}\times28)$           & 0                 & $6.0\times10^{-4}$   & $5.0\times10^{-4}$ \\
\enddata
\end{deluxetable}

\startlongtable
\begin{deluxetable}{l c l}
\tablecaption{Adopted model parameters in this population synthesis study.\label{tab:MC}}
\tablewidth{0pt}
\tablehead{
\colhead{Parameter} & \colhead{Distribution/Value} & \colhead{Description}
}
\startdata
C/O & $\mathcal{N}^{\mathrm{a}}(\mu,\sigma^{2});\ \mu=0.43,\ \sigma=0.1$ & stellar C/O ratio \\
$\mathrm{[Fe/H]}$ & $\mathcal{N}(\mu,\sigma^{2});\ \mu=0.1458,\ \sigma=0.16$ & stellar metallicity \\
$\xi_{\rm p/g}$ & $f_{\rm cdf,ice}^{\mathrm{b}}\times 10^{\mathcal{N}(\mu,\sigma^{2})}$ & pebble-to-gas flux ratio \\
$\dot{M}_{\rm g0}\ (M_\odot\,{\rm yr}^{-1})$ & $10^{\mathcal{U}^{\mathrm{c}}(-8,-7)}$ & initial disk accretion rate \\
$R_{\rm peb}\ ({\rm mm})$ & 1 & pebble radius \\
$r_{0}\ ({\rm au})$ & $\mathcal{N}(\mu,\sigma^{2});\ \mu=r_{\rm ice},\ \sigma=0.3$ & birthplace of protoplanet \\
$m_{0}\ (M_{\oplus})$ & 0.01 & birth mass of protoplanet \\
$\alpha_{\rm t}$ & $10^{\mathcal{U}(-4,-2)}$ & midplane turbulent strength \\
$\alpha_{\rm g}$ & $10^{-2}$ & global angular momentum transport coefficient \\
$C$ & 0.2 & migration reduction factor \\
\enddata
\tablenotetext{a}{$\mathcal{N}(\mu,\sigma^{2})$ denotes a normal distribution with mean $\mu$ and standard deviation $\sigma$.}
\tablenotetext{b}{$f_{\rm cdf,ice}$ is the mass fraction of solids within the ice lines of each chemical species for a solar-metallicity star; see Table~\ref{tab:chemistry}.}
\tablenotetext{c}{$\mathcal{U}(a,b)$ denotes a uniform distribution between $a$ and $b$.}
\end{deluxetable}
 
The core accretion halts once the planet reaches the pebble isolation mass $M_{\rm iso}$, the formula of which is adopted from \cite{Bitsch2018}. After that, we assume that the planet initiates substantial gas accretion, followed by the Kelvin-Helmholtz contraction, Hill-sphere limited accretion and disk flow limited accretion:   
\begin{equation}
\dot{M}_{\mathrm{p,g}}
= \min\!\left[
\left(\frac{\mathrm d M_{\mathrm{p,g}}}{\mathrm dt}\right)_{\mathrm{KH}},
\left(\frac{\mathrm d M_{\mathrm{p,g}}}{\mathrm dt}\right)_{\mathrm{Hill}},
\dot{M}_{\mathrm g}
\right].
\end{equation}
The formulas of accretion in these phases can be found in \cite{Liu2019}.

We utilize a unified planet migration torque formula derived from \cite{Kanagawa2018}:
\begin{equation}
\begin{split}
    \Gamma  &= C f_{\rm tot} \Gamma_0 \\
            &= C \left[ f_{\rm I} f_{\rm s} + f_{\rm II} \left( 1-f_{\rm s} \right) \right] \Gamma_{\rm 0},
\end{split}
\end{equation}
where $\Gamma_{\rm 0} {=} m^2 \, \Sigma \, r^4 \, \Omega_{\rm K}^2/ M_{\star}^2 \, h_{\rm g}^2$ is the normalized torque strength, $f_{\rm I}$ and $f_{\rm II}$ are the type I and type II migration coefficients. The type II migration coefficient $f_{\rm II} {=} -1$ whereas the type I migration coefficient $f_{\rm I}$ is calculated based on \cite{Paardekooper2011},  set by the disk thermal structure and local turbulent coefficient $\alpha_{\rm t}$. 
We adopt a smooth function  $f_{\rm s} = 1/[1 + (m/M_{\rm gap})^4]$ such that $\Gamma{\approx}\Gamma_{\rm I}$ at $m{\ll} M_{\rm gap}$ and $\Gamma{\approx}\Gamma_{\rm I}/(m/M_{\rm gap})^2$ at $m{\gg} M_{\rm gap}$. The gap opening mass $M_{\rm gap}$ is adopted as $2.3$ times  $M_{\rm iso}$. 
We introduce the migration reduction factor $C$ to account for the uncertainties  in the aforementioned simplified treatment of the migration process.    

We employ a Monte Carlo approach for sampling the initial conditions, including the stellar, disk and protoplanet parameters. We simulate 1000 systems for each set of Monte Carlo samples, considering the growth and migration of single protoplanets. We assume that the birth masses of the protoplanets $m_0$ are all $0.01 \ M_{\oplus}$ and their birth sites $r_0$ are associated with the ice line locations, including those for water, organic carbon, and carbon dioxide. Recent studies have proposed that near the ice line, the volatile components of pebbles evaporate. The vapor subsequently diffuses backward to the exterior of the ice line and re-condenses at lower temperatures. Such a process can enrich the local solid density near the ice line and trigger the formation of protoplanets through streaming instability \citep{2017A&A...602A..21S}. Following this scenario, a Gaussian distribution is adopted for $r_0$, with $\mu{=}r_{\rm ice}$ and $ \sigma {=} 0.3$ au. The probability of protoplanet formation near different ice lines is determined by the relative mass fraction of solid species (i.e. $p_{\rm H_2O \, ice \, line}/p_{\rm org.C \, ice \, line} {=} f_{\rm H_2O}/f_{\rm org. C}$). 

Since originated from the same molecular cloud, we consider that the metallicity of natal protoplanetary disk ($ \rm Z_{\rm d} {=} \Sigma_{\rm peb} / \Sigma_{\rm g}$) is equivalent to the metallicity of the stellar host, given by $Z_{\star} {=} Z_{\odot} \times 10^{\rm [Fe/H]}$, where $Z_{\odot} = 0.014$ or $\mathrm{[Fe/H]} = 0$ correspond to the solar metallicity. 
The mass of the host star is fixed at $1 \ M_{\odot}$ and the metallicity is randomly chosen from our observational dataset, with $\rm [Fe/H]$ following a normal distribution with $\mu {=} 0.1458$ and $\sigma {=} 0.16$. In line with observations, the stellar C/O ratio is derived from a normal distribution with a mean $\mu$ of 0.43 and a standard deviation $\sigma$ of 0.1.
The initial disk accretion rate $\dot{M}_{\rm g0}$ adheres to a log-normal distribution centered around $3 \times 10^{-8} M_{\odot} \ \rm yr^{-1}$ with a spread of 0.17 dex. The level of disk turbulence $\alpha_{\rm t}$ is selected from a log-uniform distribution spanning the range between $10^{-4}$ and $10^{-2}$. All model parameters can be found in Table \ref{tab:MC}. 

\bibliography{sample701}{}
\bibliographystyle{aasjournalv7}



\end{document}